\documentclass[twocolumn]{aastex631}

\def\gtsima{$\; \buildrel > \over\sim \;$}
\def\gtsim{\lower.5ex\hbox{\gtsima}}
\def\ltsima{$\; \buildrel < \over\sim \;$}
\def\ltsim{\lower.5ex\hbox{\ltsima}}

\received{\today}
\submitjournal{ApJ}

\shorttitle{Diagnosis of enhanced CSM with Multi Energy Neutrinos}
\shortauthors{Sawada and Ashida}

\usepackage{amsmath}
\graphicspath{{./}{figures/}}

\begin{document}

\title{Towards Multi Energy Neutrino Astronomy: \\ Diagnosing Enhanced Circumstellar Material around Stripped-Envelope Supernovae}

%\title{Novel Strategy for Diagnosing the Origin of Enhanced Circumstellar Medial around Supernovae Using Multiple MeV/TeV Neutrinos}
%\title{Diagnosis of the origin of enhanced Circumstellar Media around Type Ibc Supernova using multiple MeV/TeV-Neutrinos}

\author[0000-0003-4876-5996]{Ryo Sawada}
\altaffiliation{These two authors contributed equally to this work.}
\affiliation{Institute for Cosmic Ray Research, The University of Tokyo, Kashiwa, Chiba 277-8582, Japan}

\author[0000-0003-4136-2086]{Yosuke Ashida}
\altaffiliation{These two authors contributed equally to this work.}
\affiliation{Department of Physics, Kyoto University, Kyoto, Kyoto 606-8502, Japan}
\affiliation{Department of Physics and Astronomy, University of Utah, Salt Lake City, Utah 84112, USA}

\correspondingauthor{Ryo Sawada, Yosuke Ashida}
\email{ryo@g.ecc.u-tokyo.ac.jp; assy.8594.1207.physics@gmail.com}

\begin{abstract}
A novel approach is proposed to reveal a secret birth of enhanced circumstellar material (CSM) surrounding a collapsing massive star using neutrinos as a unique probe. 
In this scheme, non-thermal TeV-scale neutrinos produced in ejecta-CSM interactions are tied with thermal MeV neutrinos emitted from a pre-explosion burning process, based on a scenario that CSM had been formed via the pre-supernova activity.
Taking a representative model of the pre-supernova neutrinos, the spectrum and light curve of the corresponding high-energy CSM neutrinos are calculated at multiple mass-loss efficiencies, which are considered as a systematic uncertainty. 
In addition, as a part of the method demonstration, the detected event rates along time at JUNO and IceCube, as representative detectors, are estimated for the pre-supernova and CSM neutrinos, respectively, and are compared with the expected background rate at each detector. 
The presented method is found to be reasonably applicable for the range up to $\sim$1~kpc and even farther with future experimental efforts. 
The potentialities of other neutrino detectors, such as SK-Gd, Hyper-Kamiokande and KM3NeT, are also discussed. 
This is a pioneering work of performing astrophysics with neutrinos from diverse energy regimes, initiating {\it multi energy neutrino astronomy} in the forthcoming era where next-generation large-scale neutrino telescopes are operating.
\end{abstract}

\keywords{Circumstellar matter (241); Core-collapse supernovae (304); Massive stars (732); High energy astrophysics (739); Neutrino astronomy (1100); Supernova neutrinos (1666)}


%%---------------------------------------------------------------------------------
\section{Introduction} \label{sec:intro}

%%% History of neutrino astronomy 
%More than three decades have passed since mankind acquired a means of observing stars and the Universe in a probe other than light, for the first time, at an epoch-making observation of neutrinos from a supernova SN~1987A in the Large Magellanic Cloud~\citep[][]{1987PhRvL..58.1490H,1987PhRvL..58.1494B,1987JETPL..45..589A}. 
Neutrinos have brought mankind successful observations of stars via a messenger other than light over the last half century, for example from the Sun~\citep[][]{2023PhRvD.108j2005B,2024PhRvD.109i2001A} and a supernova SN 1987A~\citep[][]{1987PhRvL..58.1490H,1987PhRvL..58.1494B,1987JETPL..45..589A}.
Cosmic observation via neutrinos ---{\it neutrino astronomy}--- has been progressing by recent achievements at the IceCube Neutrino Observatory; confirmation of the astrophysical diffuse neutrino flux~\citep[e.g.,][]{2021PhRvD.104b2002A,2024PhRvD.110b2001A} as well as observations of neutrino emission from a blazar TXS~0506+056~\citep[][]{2018Sci...361.1378I} and a Seyfert galaxy NGC~1068~\citep[][]{2022Sci...378..538I}. 
One can expect this momentum to be further accelerated by more neutrino telescopes that are planned to be built in the upcoming years. 
%They are designed specifically for their own scientific purposes, producing variations in flavor and energy of detectable neutrinos. 
Each experiment is designed specifically for its own purpose, such as neutrino oscillation measurement and astrophysical neutrino observation, producing variations in flavor and energy of target neutrinos.
%For instance, liquid scintillation detectors, such as on-going KamLAND~\citep[][]{1999NuPhS..77..171S} and upcoming JUNO~\citep[][]{2022PrPNP.12303927J}, have advantages of detecting low-energy ($\lesssim$10~MeV) electron antineutrinos, while water-based Cherenkov detectors, e.g., IceCube~\citep[][]{2017JInst..12P3012A}, KM3NeT~\citep[][]{2016JPhG...43h4001A} and Super-/Hyper-Kamiokande~\citep[][]{2003NIMPA.501..418F,2018arXiv180504163H}, are better to explore high-energy (GeV--PeV) neutrinos with their gigantic size.
For instance, liquid scintillation detectors, such as on-going KamLAND~\citep[][]{1999NuPhS..77..171S} and upcoming JUNO~\citep[][]{2022PrPNP.12303927J}, are well suited for detecting low-energy ($\lesssim$20~MeV) electron antineutrinos, while water-based Cherenkov detectors are better for exploring higher energies, e.g., $\mathcal{O}(10)\,\mathrm{MeV}$--TeV at Super-Kamiokande~\citep[][]{2003NIMPA.501..418F} and Hyper-Kamiokande~\citep[][]{2018arXiv180504163H}, and TeV--PeV at IceCube~\citep[][]{2017JInst..12P3012A} and KM3NeT~\citep[][]{2016JPhG...43h4001A}.
Utilizing multiple types of detector would enable us to pursue astrophysics through detection of neutrinos from the same source over the wide energy range, as is practiced in optical surveys, i.e., open up an era of {\it multi energy neutrino astronomy}. 
The main scope of this paper is to demonstrate one example of multi energy neutrino astronomy with a special focus on circumstellar material (CSM) around core-collapsing massive stars.

%%% 
Understanding the physical properties of massive stellar outbursts is an important step towards completing stellar evolution theory \citep[e.g.,][]{2015PASA...32...15Y}.
In particular, mass-loss activity prior to a core-collapse supernova explosion (SN) is imprinted on the physical properties of the CSM, which can contribute to the radiation source of the SN \citep[see, for example,][]{2012ApJ...744...10K,2017hsn..book..875C,2017hsn..book..403S}.
Recent developments in transient surveys and rapid follow-up systems have led to an increased awareness of the diversity of the CSM structures inferred from the progenitors of SNe. 
For example, in the case of the type-Ibn SN iPTF13beo, the mass-loss rate from the progenitor Wolf-Rayet star (WR) was estimated to be about $2\times10^{-3}M_\odot \mathrm{yr}^{-1}$, far exceeding the value of about $10^{-6}M_\odot \mathrm{yr}^{-1}$ predicted by the steady-state star wind \citep[][]{2014MNRAS.443..671G}.
Standard stellar evolution theory predicts that the mass-loss rate does not increase significantly before the SN explosion \citep[][]{2012ARA&A..50..107L}, so several mechanisms have been proposed to explain the increase in the mass-loss rate that deviates from the steady-state wind structure \citep[e.g.,][]{2012MNRAS.423L..92Q, 2014ApJ...780...96S,2014ApJ...785...82S, 2017MNRAS.470.1642F,2021ApJ...906....3W,2021PASJ...73.1128T,2023ApJ...952..115T,2024ApJ...963..105M,2024A&A...685A..58E}.

%%%
One possibility is that the huge neutrino luminosity just before the stellar collapse leads to the extreme mass loss at the stellar surface \citep{2014A&A...564A..83M}. 
The mass loss from the core due to the neutrino emission leads to a sudden decrease in the gravitational force. The effect of mass loss due to high neutrino luminosities after the core collapse of a star has been established in \citet{2013ApJ...769..109L} and \citet{1980Ap&SS..69..115N}. 
\citet{2014A&A...564A..83M} suggested that for massive stars near the Eddington luminosity, the gravitational potential can be significantly reduced because of the neutrino emission shortly before the core collapse, and then the weakened gravitational force leads to the enhanced mass loss.

%%% 
Here, we propose the following strategy for testing hypotheses as a way of determining the origin of the enhanced CSM.
First, let us hypothesize that the extreme mass loss is caused by the dissipation of mass in the central core due to neutrino emission $L_{\mathrm{pre}\mathchar`-\nu}$ just before the explosion of a massive star \citep[predicts the CSM density $\rho_\mathrm{csm}\propto L_{\mathrm{pre}\mathchar`-\nu}$;][]{2014A&A...564A..83M}.
In the central core of a massive star after the carbon-burning stage, very-high-luminosity thermal neutrino emission $(\mathcal{O}(1)\,\mathrm{MeV})$ is expected, and if the star is close enough to Earth, its neutrinos can be detected by the latest neutrino detectors several weeks before collapse of the core \citep[][]{2017ApJ...851....6P}.
Furthermore, it has also been proposed that the high-density CSM formed by the extreme mass loss may interact with the SN shock wave to produce high-energy neutrino emission \citep[predicts $L_{\mathrm{CSM}\mathchar`-\nu}\propto\rho_\mathrm{csm}$; ][]{2018PhRvD..97h1301M,2024PhRvD.109j3020M}.
It has also been shown that a high-statistics $\mathcal{O}(10)\,\mathrm{TeV}$ neutrino signal in IceCube is expected even for an ordinary SN in the Galaxy. 
Therefore, we can hypothesize that if the dense CSM around a SN progenitor is originally caused by release of pre-SN neutrinos, then there should be characteristic correlation between pre-SN neutrinos and high-energy neutrinos from the same astronomical object ($L_{\mathrm{pre}\mathchar`-\nu}\propto L_{\mathrm{CSM}\mathchar`-\nu}$ in a broad sense). 

%<<<
\begin{figure*}[htbp]
    \centering
    \includegraphics[width=0.95\textwidth]{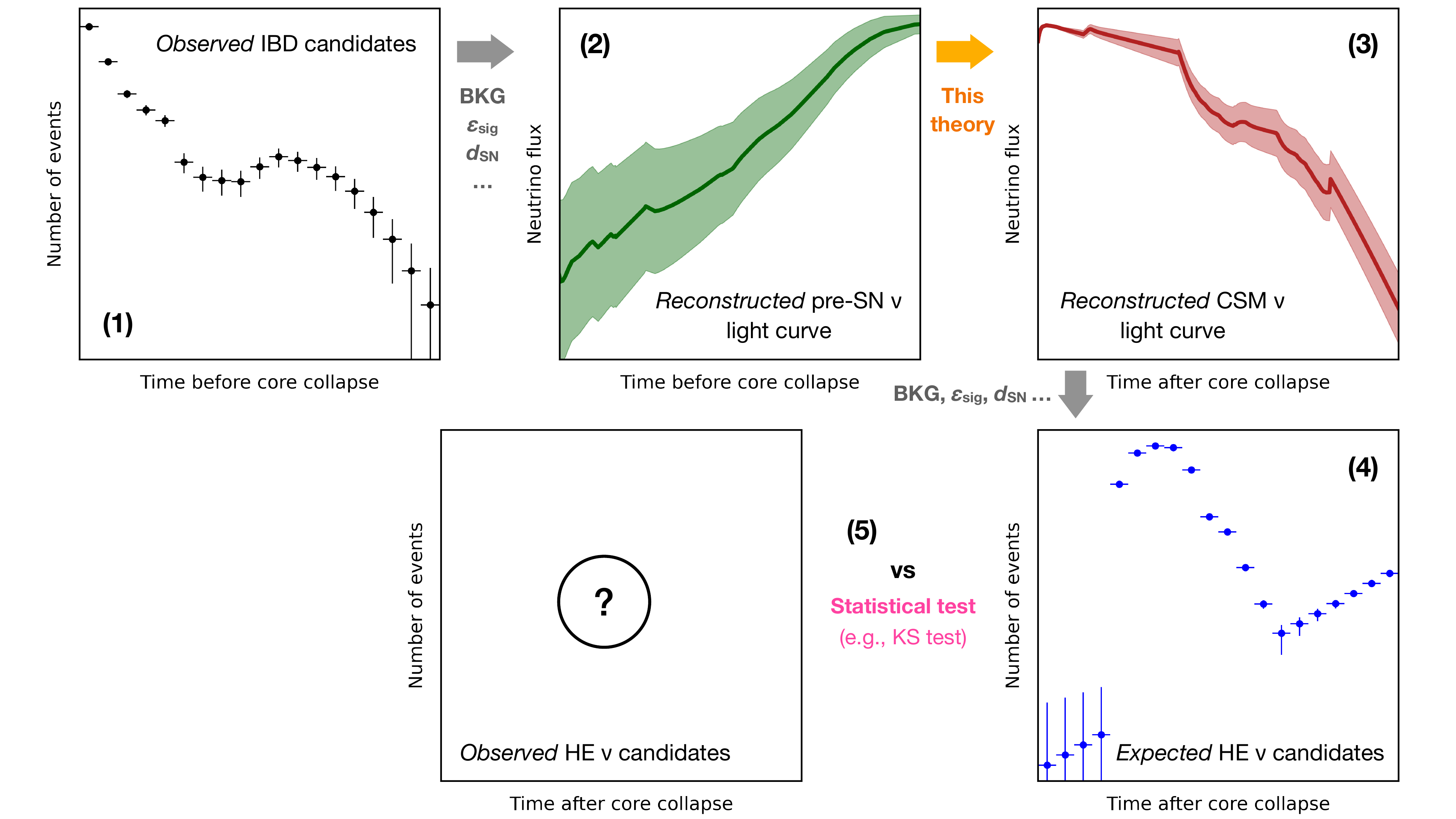}
    \vspace{+5truept}
    \caption{Schematic diagram for the proposed flow of diagnosing the CSM origin with low- and high-energy neutrino detections. When reconstructing the pre-SN neutrino light curve from the observed inverse beta decay events at a detector, systematic uncertainties regarding estimated background, signal detection efficiency, distance estimation, and so on, as well as statistical uncertainty should be considered. This is same for when calculating the expected number of high-energy neutrino events at another detector. The proposed method can be practiced in a self-consistent way with requiring no specific pre-SN neutrino model.}
\label{fig:flow}
\end{figure*}
%>>> 

%%% Intro to the diagnosis method together with short descriptions on each section 
In the present work, we propose a novel method to study the origin of the enhanced CSM by detecting low- and high-energy neutrinos at relevant detectors (JUNO and IceCube are mainly considered in this paper). 
Our methodology is schematically illustrated in Figure~\ref{fig:flow}; each step is briefly introduced below, with further details provided in the corresponding section. 
\begin{enumerate}
    \item[(1)] Select low-energy neutrino event candidates at a detector and capture the detected number in time. 
    \item[(2)] Reconstruct pre-SN neutrino light curve based on the selected event number, expected background, signal detection efficiency, estimated distance to the SN, etc. 
    \item[(3)] Reconstruct non-thermal high-energy neutrino light curve using the presented model in this paper. 
    \item[(4)] Calculate the high-energy neutrino event rate in time expected at a detector using neutrino cross section, detection efficiency, background, distance to the SN, etc. 
    \item[(5)] Perform a statistical test to compare the expected and observed time distributions. 
\end{enumerate}
The theoretical model used in the procedure (3) is explained in Sec.~\ref{sec:model}, and the other procedures (1), (2), (4), and (5), bridging between theory and experiment, are described in Sec.~\ref{sec:demo}.
Concluding remarks and discussion are given in Sec.~\ref{sec:concl}.

%%%%
%When estimating the high-energy neutrino event rate, depending on the supernova direction, we should consider the effect of neutrino interactions inside the Earth. 
%In the KS test, systematic uncertainties regarding experiments, distance estimation, as well as statistical uncertainty should be taken into account. 
%The proposed method is self-consistent with requiring no specific pre-SN neutrino model. 
%Only a thing we assume in the above procedure is the model for converting the pre-SN neutrino light curve to non-thermal high-energy neutrino light curve, as described in Section~\ref{sec:model}. 
%, e.g., Kolmogorov–Smirnov (KS) test

%%---------------------------------------------------------------------------------
\section{Model} \label{sec:model}

In the theoretical model provided in this study, we first assume a pre-SN neutrino light curve model (Sec. \ref{sec:mm_presn}), and then reconstruct the CSM density profiles from this (Sec. \ref{sec:mm_csm}).
Given the shock evolution (Sec. \ref{sec:mm_sn}), we predict the non-thermal high-energy neutrino light curve  (Sec. \ref{sec:mm_hinu}).

%%%
\subsection{Pre-SN Neutrino Model} \label{sec:mm_presn} 
We use the 15$M_\odot$ progenitor model of \citet{2015ApJ...808..168K,2017ApJ...848...48K} 
for the neutrino luminosities $L_{\mathrm{pre}\mathchar`-\nu}$ of massive stars shortly before the core collapse, as shown in Figure \ref{fig:elumi} (a--1).
In addition, the neutrino energy spectra are shown in Figure \ref{fig:elumi} (a--2) and (a--3).
Many previous studies have shown that the precursor neutrino luminosities are roughly the same across the typical SN progenitor mass range \citep[e.g.,][]{Odrzywolek:2010zz,2016PhRvD..93l3012Y,2017ApJ...851....6P,2020ARNPS..70..121K}.

%%%
\subsection{Reconstructing the CSM from Pre-SN $\nu$} \label{sec:mm_csm}
In this study, the stellar mass-loss rate $\dot{M}_\star$ is assumed to be enhanced in response to the core mass-loss rate $\dot{M}_\mathrm{c}$ \citep{2014A&A...564A..83M}. 
The core mass-loss rate $\dot{M}_\mathrm{c}$ with neutrino luminosity $L_\nu$ can be estimated from a simple relation $L_\nu=\dot{M}_\mathrm{c}c^2$, as shown in Figure~\ref{fig:elumi}.
In order to take into account the effect of the core mass loss being partially relaxed inside the star, we introduce the mass-loss efficiency $\beta$ between the core mass loss $\dot{M}_\mathrm{c}$ and the mass-loss rate from the stellar surface $\dot{M}_\star$ as follows
\begin{align}
    \dot{M}_\star(t) 
    &\approx \dot{M}_{wind}+\beta\dot{M}_\mathrm{c}(t) 
    \nonumber \\
    &= \dot{M}_{wind}+\beta\cdot\frac{L_{\mathrm{pre}\mathchar`-\nu}(t)}{c^2}~,\label{eq:pre-nu}
\end{align}
where $\dot{M}_{wind}$ is the steady-state wind mass-loss rate, and we fix it to $10^{-6}M_\odot\,\mathrm{yr}^{-1}$.
Here, the predicted mass-loss rate at the final stage of stellar evolution is estimated to be $10^{-7}-10^{-5} M_\odot\,\mathrm{yr}^{-1}$ \citep[][]{2012ARA&A..50..107L}.
Note that, as shown on the right axis of Figure \ref{fig:elumi} (a--1), the core mass-loss rate is $\dot{M}_\mathrm{c}\approx10^{-2}M_\odot\,\mathrm{yr}^{-1}$ at $t=10^5$ sec before the explosion.
Even if $\beta=0.01$ is adopted, $\dot{M}_\mathrm{c}\gg\dot{M}_{wind}$, suggesting that the contribution of the steady wind $\dot{M}_{wind}$ is small.

Next, we adopt the framework for describing the CSM structure derived from the time-dependent mass-loss history presented in \cite{2020ApJ...894....2P}.
We assume that the ejection velocity remains constant at the escape velocity of the stellar surface $v_\mathrm{csm}=v_\mathrm{esc}=\sqrt{2GM_\star/R_\star}$. 
When an ejection is launched from a stellar surface $R_\star$ at time $t_\mathrm{csm}$, then $t_\mathrm{csm}$ can be represented by the current position $r$ and time $t$ as
\begin{equation}
    t_\mathrm{csm} 
    = t - \frac{r-R_\star}{v_\mathrm{csm}}~.
\end{equation}
This provides the density profile of the CSM $\rho_\mathrm{csm}(r,t)$ at any time $t$ as
\begin{equation}    
    \rho_\mathrm{csm}(r,t) = \frac{\dot{M}_\star(t_\mathrm{csm} )}{4\pi r^2\cdot v_\mathrm{csm}}~.
\end{equation}
Since this process is expected to operate in Wolf-Rayet stars \citep{2014A&A...564A..83M}, we adopt $M_\star=5M_\odot$ and $R_\star=3\times10^{11}\,\mathrm{cm}$ as a progenitor profile for this study.
If the pre-SN neutrino luminosity is sufficiently high (i.e. $\dot{M}_{wind}\ll\beta\dot{M}_\mathrm{c}(t)$), the density $\rho_\mathrm{csm}(r,t)$ scales roughly as $\rho_\mathrm{csm}(r,t)\propto \beta\cdot L_{\mathrm{pre}\mathchar`-\nu}(t_\mathrm{csm})/r^2$.

%%%
\subsection{Shock Evolution} \label{sec:mm_sn}
The shocked high-density CSM and SN ejecta form a thin shell because of efficient radiative cooling. 
For the evolution of the shock velocity, we adopt the well-known thin-shell model for the ejecta \citep{1992ApJ...388..103K} that has been used in \citet{2018PhRvD..97h1301M,2024PhRvD.109j3020M}.

Assuming that the shocked region is geometrically thin and hence its location,
velocity, and mass are given by $R_\mathrm{sh}$, $v_\mathrm{sh}$, and $M_\mathrm{sh}$, the time evolution of the shell is described by two equations
\begin{align}
    M_\mathrm{sh}\frac{dv_\mathrm{sh}}{dt}&=4\pi R_\mathrm{sh}^2 \left[\rho_\mathrm{ej}(v_\mathrm{ej}-v_\mathrm{sh})^2 - \rho_\mathrm{csm}(v_\mathrm{sh}-v_\mathrm{csm})^2 \right]~, \label{eq:conv1}\\
    \frac{dM_\mathrm{sh}}{dt}&=4\pi R_\mathrm{sh}^2 \left[\rho_\mathrm{ej}(v_\mathrm{ej}-v_\mathrm{sh}) - \rho_\mathrm{csm}(v_\mathrm{sh}-v_\mathrm{csm}) \right]~,  \label{eq:conv2}
\end{align}
where $\rho_\mathrm{ej}$ and $v_\mathrm{ej}=R_\mathrm{sh}/t$ are the density and velocity of the unshocked SN ejecta at $R_\mathrm{sh}$, and $\rho_\mathrm{csm}$ and $v_\mathrm{csm}$ are the unshocked density and velocity, respectively.
For simplicity, we do not explicitly incorporate radiative cooling in these equations. 
We further assume that the unshocked SN ejecta expand homologously $(r=v\cdot t)$, and we employ a double power-law density structure (Eqs. \ref{eq:snprofile1} and \ref{eq:snprofile2}) based on numerical simulations
of SN explosions \citep[e.g.,][]{1994ApJ...420..268C,1999ApJ...510..379M}. 

%%% 
\subsection{High-energy SN Neutrinos} \label{sec:mm_hinu}
There is neutrino emission not only from thermal processes but also from non-thermal mechanisms.
When a supernova explodes, a collisionless shock wave mediated by plasma instability is formed. 
In such a region, charged particles such as protons are accelerated to high energies by diffusive shock acceleration (DSA), producing mesons via inelastic {\it pp} collisions.
These mesons subsequently decay into high-energy neutrinos, e.g., $pp \rightarrow \pi^{+} \rightarrow \mu^{+}\nu_{\mu} \rightarrow e^{+}\nu_{e}\nu_{\mu}\bar{\nu}_{\mu}$.
\cite{2018PhRvD..97h1301M,2024PhRvD.109j3020M} demonstrated that such non-thermal neutrino emission can occur over timescales of hours to years after explosion, with energies spanning $10^{-3}$ to $10^3$~TeV. 
Our calculations follow the neutrino production methodology of \citet{2018PhRvD..97h1301M,2024PhRvD.109j3020M} and \citet{2024arXiv240918935K}.

%<<<
\begin{figure*}[htbp]
    \gridline{\fig{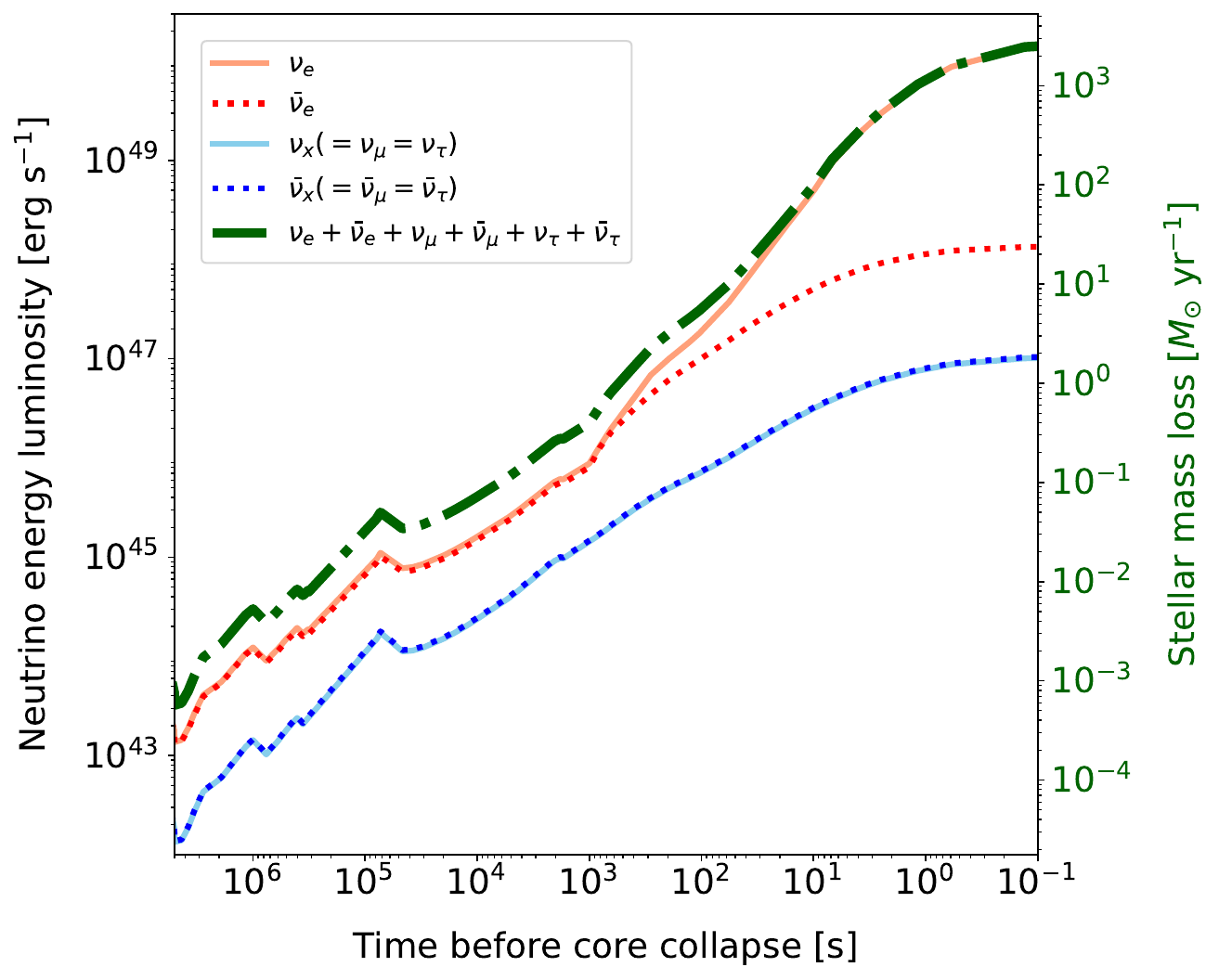}{0.46\textwidth}{(a--1) pre-SN $\nu$, no oscillation}\hspace{-30truept}
            \fig{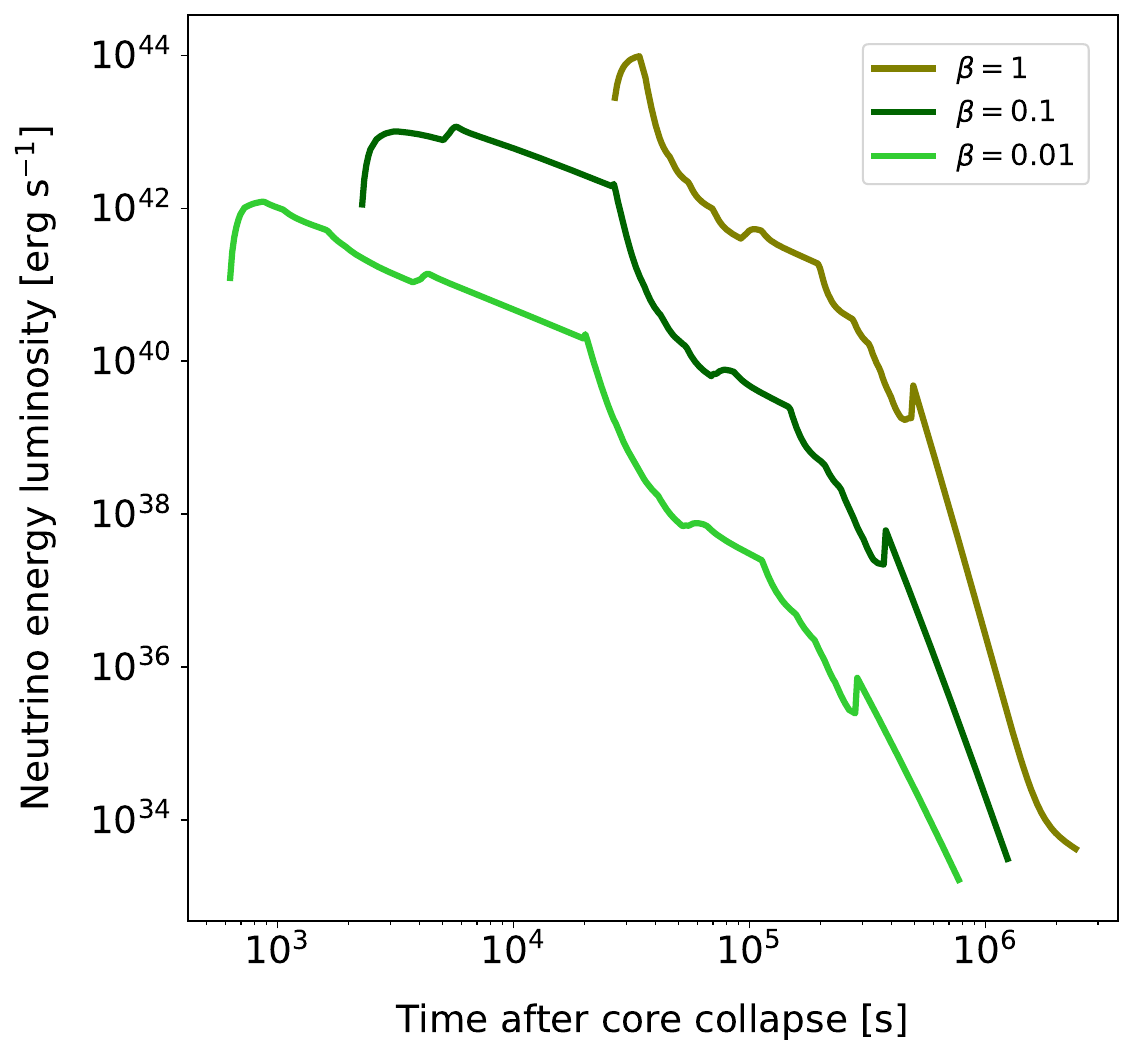}{0.4\textwidth}{(b--1) CSM $\nu$}}
    \gridline{\fig{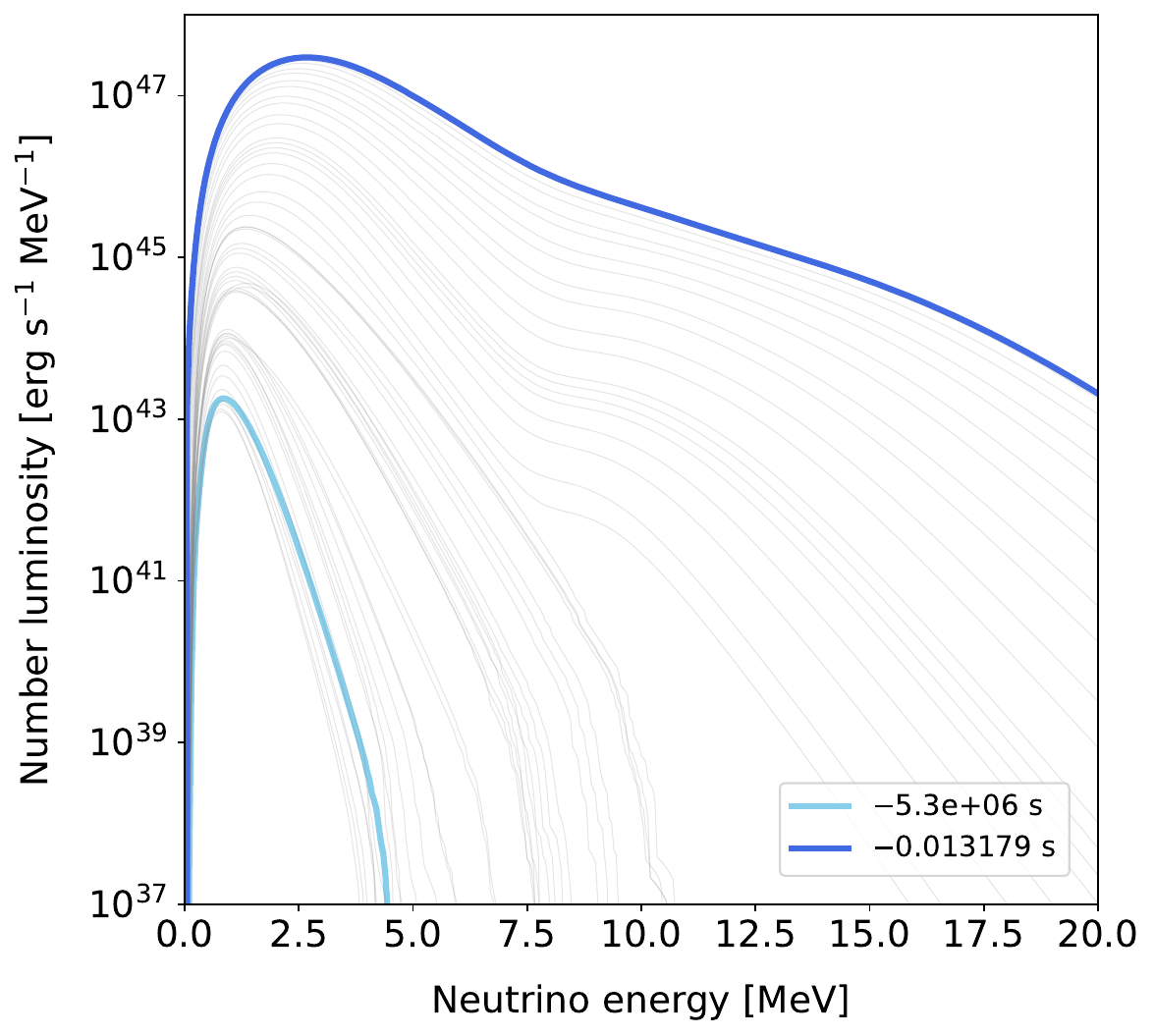}{0.4\textwidth}{(a--2) pre-SN $\nu$, NH}\hspace{-15truept}
            \fig{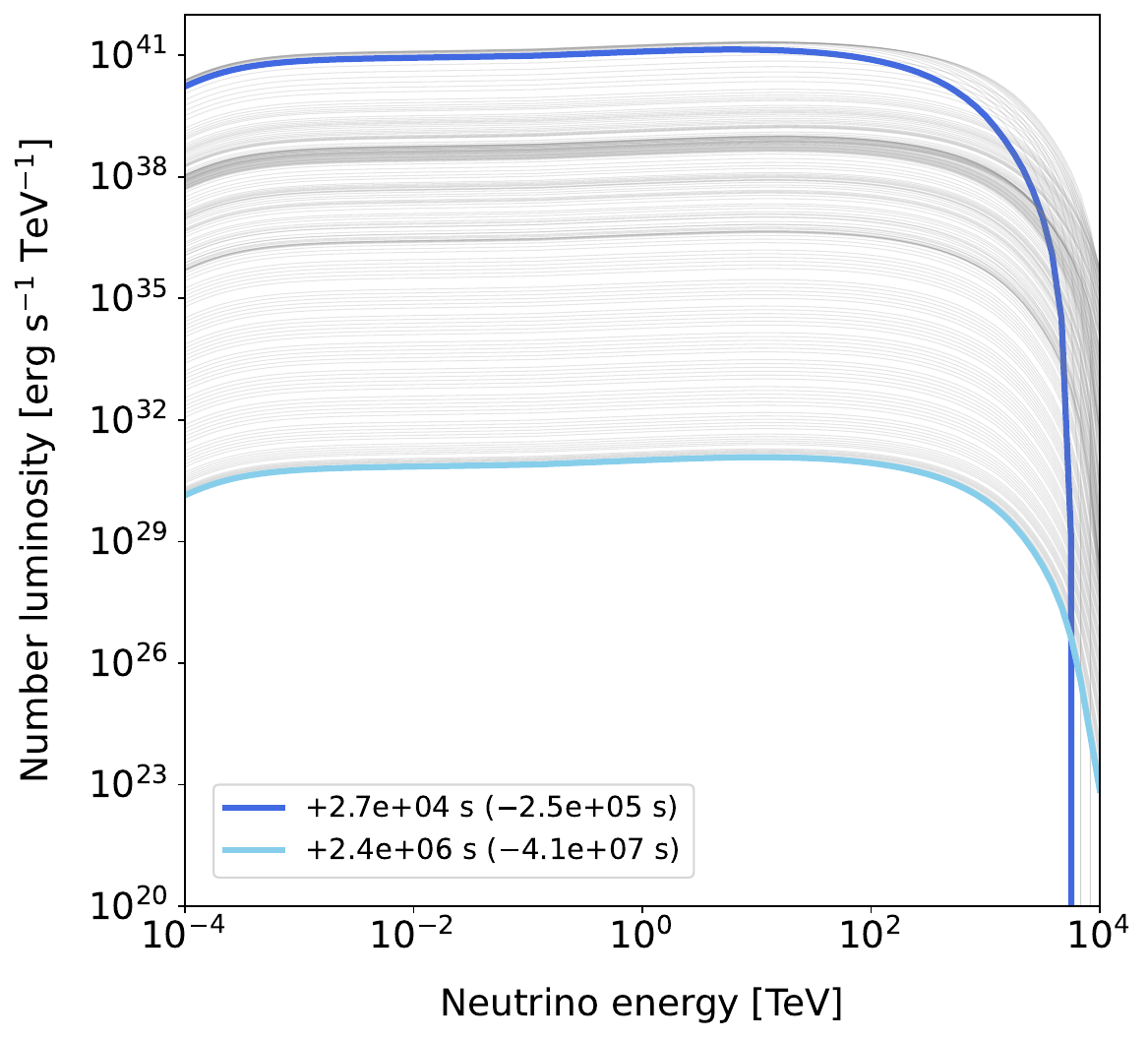}{0.4\textwidth}{(b--2) CSM $\nu$, $\beta=1$}}
    \gridline{\fig{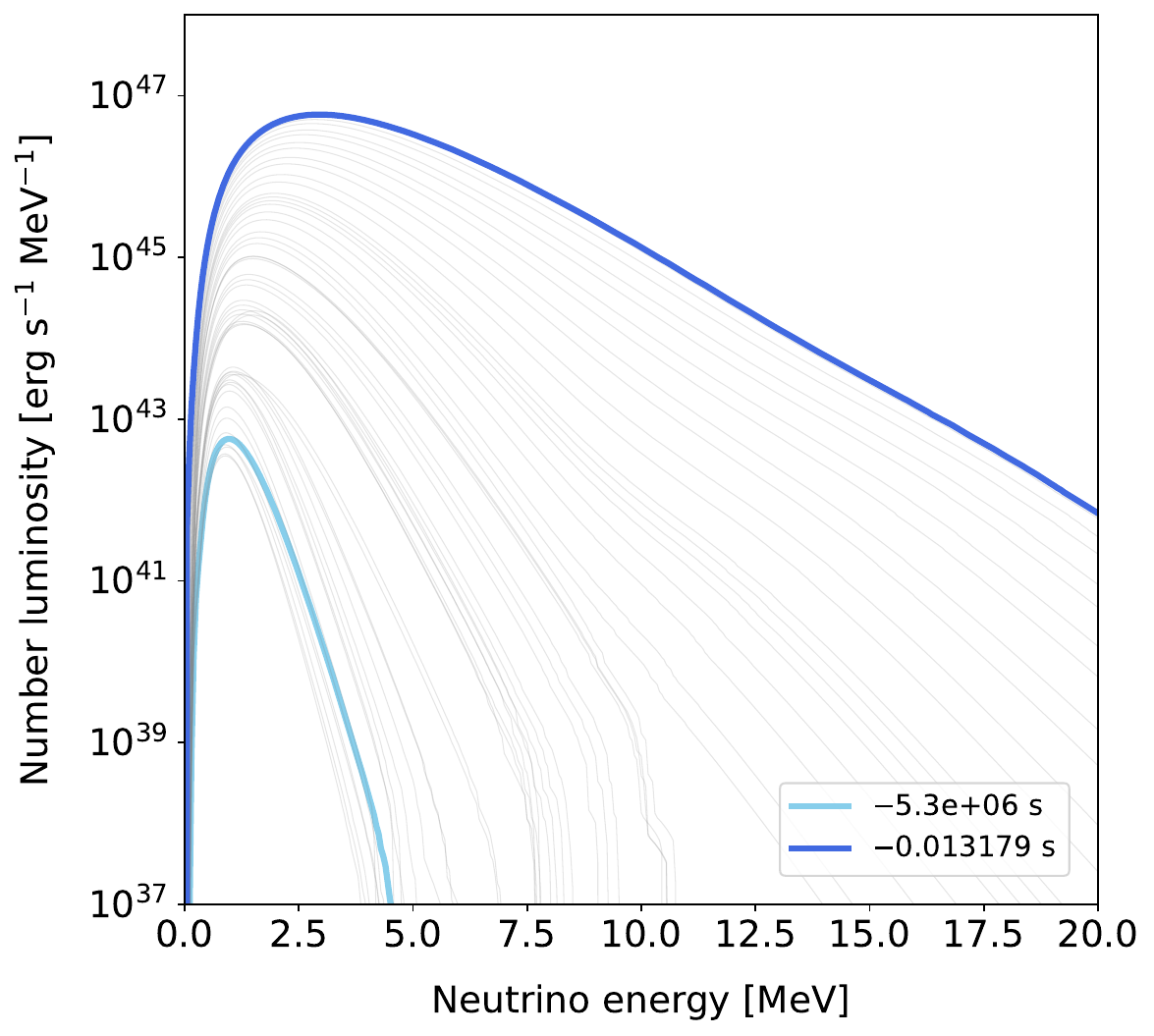}{0.4\textwidth}{(a--3) pre-SN $\nu$, IH}\hspace{-15truept}
            \fig{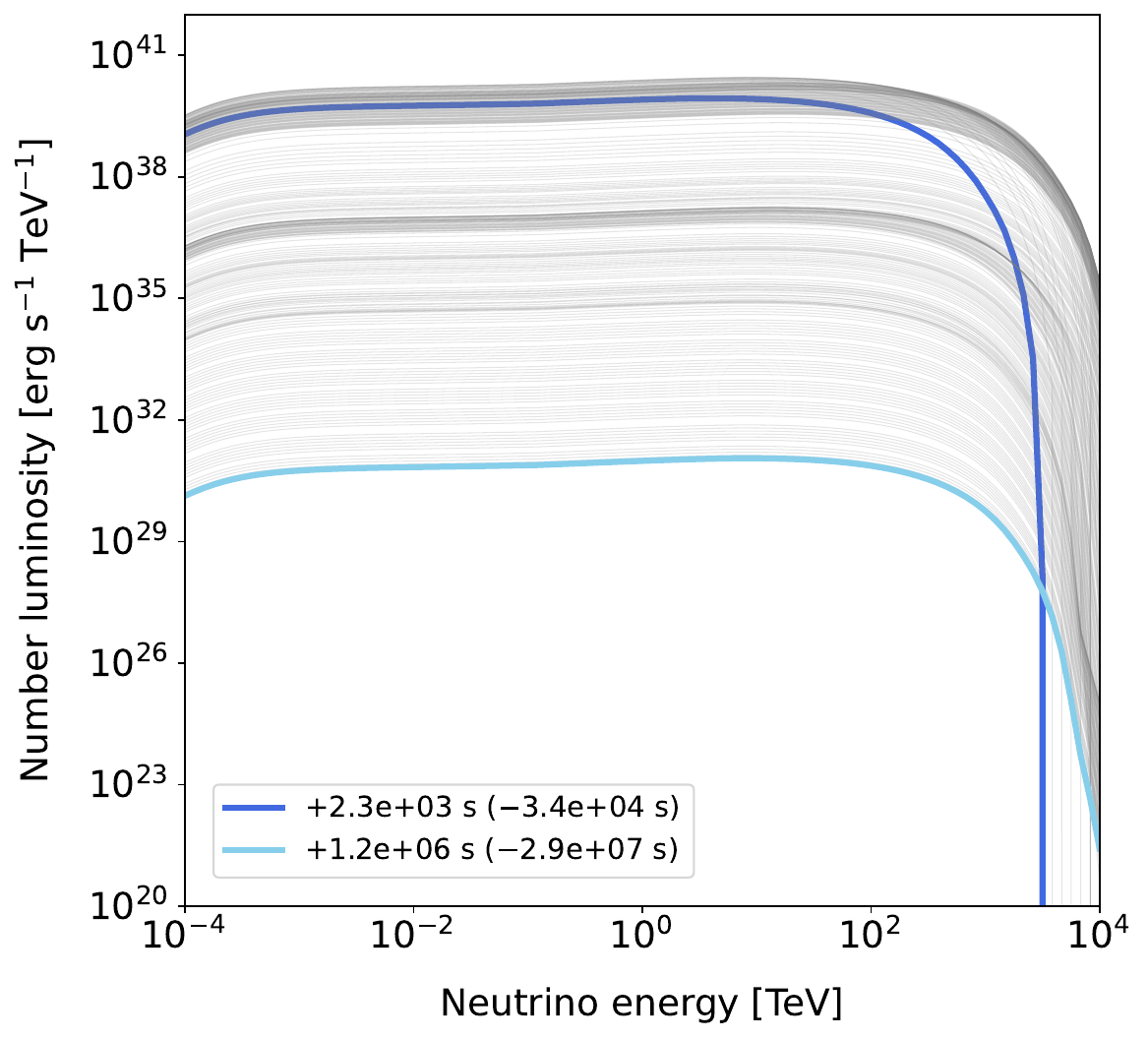}{0.4\textwidth}{(b--3) CSM $\nu$, $\beta=0.1$}}    %\vspace{-10truept}
    \caption{In the top two panels, the total energy luminosity as a function of time before and after core collapse is shown for (a--1) pre-SN neutrinos and (b--1) CSM neutrinos (all flavor sum), respectively. In the panel (a--1), the stellar mass loss is converted from the total luminosity via $L_\nu=\dot{M}_\mathrm{c}c^2$. In the other four panels, the number luminosities at different time slices of pre-SN electron antineutrinos with the mass hierarchy (a--2) NH and (a--3) IH and of CSM neutrinos (all flavor sum) with the mass-loss efficiency (b--2) $\beta=1$ and (b--3) $\beta=0.1$ are given. Here, the gray lines represent curves between the two blue ones in time.}
\label{fig:elumi}
\end{figure*}
%>>>

The `onset' time of acceleration corresponds to the formation conditions of the collisionless shock wave \citep{2018PhRvD..97h1301M}.
When the CSM is shallow, the formation of the collisionless shock wave is guaranteed immediately after the time $t_\star$ when the SN shock wave leaves a star, for which the condition is given by $R_\mathrm{sh}(t_\star)=R_\star$.
On the other hand, when the CSM is dense, the shock is initially radiative, and the condition of the collisionless shock wave formation is given by the photon escape time $t_\mathrm{bo}$; 
the timing at which $\tau_\mathrm{T}\approx\sigma_\mathrm{T}\rho_\mathrm{csm}R_\mathrm{sh}/(\mu_e m_\mathrm{H})\lesssim c/V_\mathrm{sh}$ is satisfied, where $\tau_\mathrm{T}$ is the optical depth of Thomson scattering with a cross section $\sigma_\mathrm{T}\approx6.7\times10^{-25}\mathrm{cm}^2$.
Considering these, the onset time of acceleration is given by
\begin{equation}
    t_\mathrm{onset} \approx \mathrm{max}[t_\mathrm{\star},\,t_\mathrm{bo}]~.\label{eq:onset}
\end{equation}

Shock dissipation converts the kinetic energy into heat, magnetic fields, and cosmic rays (CRs).
We assume a power-law CR spectrum as
\begin{equation}
    \frac{dn_\mathrm{CR}}{dE_p} \propto E^{-s}\exp{(-E/E^\mathrm{max}_p)}~,
\end{equation}
where the maximum proton energy, $E^\mathrm{max}_p$, is limited to the range where the CR acceleration time \citep[$t_\mathrm{acc}\approx(20/3)cE_p/(eBV_\mathrm{s}^2)$;][]{1983RPPh...46..973D} is shorter than the particle escape time ($t_\mathrm{esc} \approx R_\mathrm{sh}/c$).
We set $s = 2$ throughout this paper for simplicity.
The CR spectrum is normalized by the CR energy density,
\begin{equation}
    u_\mathrm{CR}
    =\int dE_p \, \frac{dn_\mathrm{CR}}{dE_p}~,
    \label{eq:}     
\end{equation}
where we parameterize as $u_\mathrm{CR}= \epsilon_{p} \cdot \rho_\mathrm{csm} v_\mathrm{sh}^2/2$.
Following \cite{2014ApJ...783...91C}, we adopt $\epsilon_{p}=0.1$ as a reference value.
We also parameterize the magnetic field in the emission region via $u_\mathrm{B}  = {B^2}/{(8\pi)} = \epsilon_{B} \cdot \mathcal{E}_\mathrm{expl} /\mathcal{V}_\mathrm{sh}$, where $\mathcal{V}_\mathrm{sh}\approx(4\pi/3)R_\mathrm{sh}^3$ is the volume, $\mathcal{E}_\mathrm{expl}$ is the SN explosion energy, and $\epsilon_{B}=0.01$ is used in this work.

The neutrino production rate $ \Phi(E_\nu) dE_\nu$ in the energy interval $(E_\nu, E_\nu+ dE_\nu)$ per unit hydrogen number density can be written as
\begin{equation}
    \Phi(E_\nu) = \int_{E_{p,\mathrm{sim}}}^{\infty} 
    \sigma_{pp}(E_p)\cdot  
    F_\nu(E_\nu, E_p) \cdot 
    n_\mathrm{CR}(E_p) \frac{dE_p}{E_p} ~,
\end{equation}
where $F_\nu(E_\nu, E_p)$ is the differential spectrum of the secondary particles, and its parametrization follows Eqs. (62) and (66) of \citet{2006PhRvD..74c4018K}.  
The function $\sigma_{pp}(E_p)$ is the inelastic cross section of $pp$ interactions, following the post-Large-Hadron-Collider formula given by \citet{2014PhRvD..90l3014K}.

The differential energy injection rate of neutrinos in a time-dependent manner is then given by
\begin{equation}
    \dot{n}^\mathrm{inj}_\nu(E_\nu) = \frac{cM_\mathrm{CSM}}{m_\mathrm{H}\mathcal{V}_\mathrm{sh}} \cdot \Phi(E_\nu)~.
    \label{eq:}     
\end{equation}
We solve the neutrino transport equation
\begin{equation}
    \frac{\partial}{\partial t} n_\nu(E_\nu)= - \frac{n_\nu(E_\nu)}{t_\mathrm{esc}}+ \dot{n}^\mathrm{inj}_\nu(E_\nu)~.
    \label{eq:}     
\end{equation}
The differential neutrino luminosity [erg s$^{-1}$] is calculated by
\begin{equation}
    L_\nu(E_\nu) = \frac{E_\nu^2 \cdot n_{\nu}(E_\nu) \cdot \mathcal{V}_\mathrm{sh}}{t_\mathrm{esc}}~.
    \label{eq:}     
\end{equation}

Figure \ref{fig:elumi} (b--1) shows the total energy luminosity of CSM neutrinos as a function of time after the core collapse.
In addition, the neutrino energy spectra are shown in Figure \ref{fig:elumi} (b--2) and (b--3).
Notably, the CSM neutrino emission begins at later times when the mass-loss efficiency $\beta$ is higher. 
Even under the same pre-SN neutrino luminosity, a larger $\beta$ results in a denser CSM $\rho_\mathrm{csm}$ ($\propto \beta\cdot L_{\mathrm{pre}\mathchar`-\nu}$), delaying the onset time of acceleration as given by Eq.~\eqref{eq:onset}.

%%---------------------------------------------------------------------------------
\section{Demonstration} \label{sec:demo}

In this section, we demonstrate the proposed method based on the number of expected events of pre-SN and CSM neutrinos at relevant detectors. 
We consider primarily JUNO and IceCube as a detector for pre-SN and CSM neutrinos, respectively, while discussion about potentialities of other detectors is also given later. 

%%% 
\subsection{Pre-SN Neutrino Detection} \label{sec:deteclowe}

In the MeV energy regime, the main detection channel is usually inverse beta decay (IBD) of electron antineutrinos ($\bar{\nu}_e + p \rightarrow e^+ + n$) as its cross section overwhelms other channels at these energies. 
JUNO is a 20-kton liquid scintillation detector under construction in China~\citep{2022PrPNP.12303927J}. 
It has multiple purposes of physics program, including SN neutrinos, reactor neutrinos as well as proton decay search. 
At JUNO, a delayed coincidence (DC) strategy is used to identify IBD events, where scintillation photons from the primary positron and the subsequent 2.2-MeV $\gamma$-ray from thermal neutron capture on hydrogen are detected by the instrumented photomultiplier tubes (PMTs). 
Requiring DC helps to reduce a large amount of low-energy background events that exist in the region of current interest, e.g., PMT noise and ambient neutrons. 
A previous study performed using a dedicated Monte Carlo simulation shows the event rates for the 1.8--4.0~MeV range are, 15.7~day$^{-1}$ reactor neutrinos, 1.1~day$^{-1}$ geo neutrinos, $<$0.9~day$^{-1}$ accidental coincidence, $<$0.1~day$^{-1}$ fast neutrons, and $<$0.1~day$^{-1}$ $(\alpha, n)$, with an IBD detection efficiency of 72.9\% after a series of analysis cuts~\citep[][]{2016JPhG...43c0401A}. 
In total, the expected background rate in the pre-SN neutrino detection is 18.0~day$^{-1}$.

One can estimate the pre-SN neutrino event rate at JUNO as 
\begin{equation}
    \frac{dN_{\rm event}}{dt} = \frac{1}{4 \pi d^2} \int_{E_{\rm min}}^{E_{\rm max}} \Phi_{\rm pre\mathchar`-\nu}(E_\nu) \cdot \sigma_{\rm IBD}(E_\nu) \cdot N_{\rm t} \cdot \epsilon_{\rm sig} dE_\nu~, 
\end{equation}
where $\Phi_{\rm pre\mathchar`-\nu}$ is the pre-SN $\bar{\nu}_e$ flux at a source [s$^{-1}$], $\sigma_{\rm IBD}(E_\nu)$ is the $\bar{\nu}_e$-$p$ IBD cross section [${\rm cm^2}$] calculated by \citet{2003PhLB..564...42S}, $N_{\rm t} = 1.45 \times 10^{33}$ is the number of target free protons in JUNO (20~kton), $d$ is distance to the SN, and $\epsilon_{\rm sig} = 0.729$ is the IBD selection efficiency by the analysis proposed in \citet{2016JPhG...43c0401A} that is assumed to be constant over the current energies. 
We take $E_{\rm min} = 1.8$~MeV and $E_{\rm max} = 4.0$~MeV in the current study as this choice gives the well-estimated background rate available in \citet{2016JPhG...43c0401A}. 

The estimated number of pre-SN neutrino signals based on \citet{2017ApJ...848...48K} is shown together with background events in the left panels of Figure~\ref{fig:eventhist}. 
Here, we show both neutrino mass hierarchy cases, normal (NH) and inverted (IH), and two different distances (300~pc and 1~kpc). 
The detectable time range is determined by requiring $N_{\rm sig}/\sqrt{N_{\rm sig}+N_{\rm bkg}}>2$ and $N_{\rm sig}>1$ in this study, but this can be expanded by the potential background reduction and analysis improvements. 
In the current setup, pre-SN neutrinos can be well detected from $\sim$1 week prior to core collapse from the SN at 300~pc away in both NH and IH cases. 
It is found the detectable range is limited to less than $\sim$1~kpc because of statistics and background rate, as shown in the left bottom panel of Figure~\ref{fig:eventhist}. 
However, it should be noted that this can improve by many factors as discussed in Sec.~\ref{sec:concl}. 

%%% 
\subsection{CSM Neutrino Detection} \label{sec:detechighe}

At high energies above a few hundred GeV, neutrinos interact deep inelastically with quarks in nucleons. 
Charged-current (CC) deep inelastic (DIS) interactions produce a charged lepton and a hadronic shower in the final state, while neutral-current (NC) DIS interactions leave only hadronic cascades. 
In addition, electron antineutrinos have an enhanced interaction with an electron around 6.3~PeV that produces an on-shell $W$ boson, as known as the Glashow resonance (GR)~\citep[][]{1960PhRv..118..316G}.
Such DIS and GR events can be detected directly via Cherenkov photons by IceCube. 
In the TeV--PeV energy region, background is primarily composed of atmospheric neutrinos and astrophysical neutrinos. 
We refer to the HAKKM~2014 calculation~\citep[][]{2015PhRvD..92b3004H} for atmospheric neutrinos and to the recent IceCube measurement result with starting events~\citep[][]{2024PhRvD.110b2001A} for astrophysical neutrinos. 
We use the effective area $\mathcal{A}_{\rm eff}$ released by IceCube~\citep[][]{2021PhRvD.104b2002A} for the event rate estimation. 
It includes effects of the neutrino absorption by Earth due to large interaction cross sections as well as selection efficiency by the analysis, and differs for neutrino flavors and energies. 
Here, the neutrino-nucleon cross sections from \citet{2011JHEP...08..042C} and \citet{1960PhRv..118..316G} are considered.
Although the effective area should depend on the zenith angle because of the different Earth absorption effect, the referred one is averaged over angles, which may lead to an overestimate or underestimate of both signal and background depending on the SN direction with respect to the detector. 
This is, however, not a critical impact on our conclusion, because the background rate is much smaller than CSM neutrinos in the region of present interest as shown below. 
Another notice is raised about atmospheric neutrinos; the referred atmospheric flux is averaged over zenith angles, and the passing fraction, or self-veto effect~\citep[][]{2018JCAP...07..047A,2021PhRvD.104b2002A}, which is not taken into account in the referred effective area, is not reflected in the background estimation. 
Note again that this treatment should not change the current scope for the same reason as above.

Event rate at IceCube of CSM neutrinos can be estimated by 
\begin{equation}
    \frac{dN_{\rm event}}{dt} = \frac{1}{4 \pi d^2} \int_{E_{\rm min}}^{E_{\rm max}} \Phi_{\rm CSM\mathchar`-\nu}(E_\nu) \cdot \mathcal{A}_{\rm eff}(E_\nu) dE_\nu~, 
\end{equation}
where $\Phi_{\rm CSM\mathchar`-\nu}(E_\nu)$ is the CSM neutrino light curve at a source [s$^{-1}$], $\mathcal{A}_{\rm eff}(E_\nu)$ is the effective area taken from \citet{2021PhRvD.104b2002A} [m$^2$], and $d$ is distance to the SN. 
We set the analysis window as $E_{\rm min} = 10$~TeV and $E_{\rm max} = 1000$~TeV. 
In contrast, the atmospheric and astrophysical background estimate is performed as 
\begin{equation}
    \frac{dN_{\rm event}}{dt} = 4 \pi \int_{E_{\rm min}}^{E_{\rm max}} (\Phi_{\rm atmos\mathchar`-\nu}+\Phi_{\rm astro\mathchar`-\nu}) \cdot \mathcal{A}_{\rm eff} dE_\nu~, 
\end{equation}
where $\Phi_{\rm atmos\mathchar`-\nu}(E_\nu)$ and $\Phi_{\rm astro\mathchar`-\nu}(E_\nu)$ are atmospheric and astrophysical neutrinos, respectively [m$^{-2}$ s$^{-1}$ sr$^{-1}$], and $4 \pi$ stands for the integration over solid angles. 
The resulting event rate histograms are given in the right panels of Figure~\ref{fig:eventhist} for different mass-loss efficiencies ($\beta = 1, 0.1, 0.01$) and distances ($d = 300$~pc, 1~kpc). 
The signal event rate is overwhelmingly larger than the expected background rate in any case. 
%Notably, the result indicates that the CSM neutrino emission begins at later times for the case of the higher mass-loss efficiency. This is explained by the fact that the larger $\beta$, even for the same pre-SN neutrino luminosity, results in the higher CSM density $\rho_\mathrm{csm}$ ($\propto \beta\cdot L_{\mathrm{pre}\mathchar`-\nu}$), leading to the later onset time of acceleration as given by Eq.~\eqref{eq:onset}.

%%%
%\subsection{\memo{The available parameter space for our strategy}} \label{sec:detechighe}
\subsection{Synchronized Detection} \label{sec:synchro}

\begin{figure*}[htbp]
    \gridline{\fig{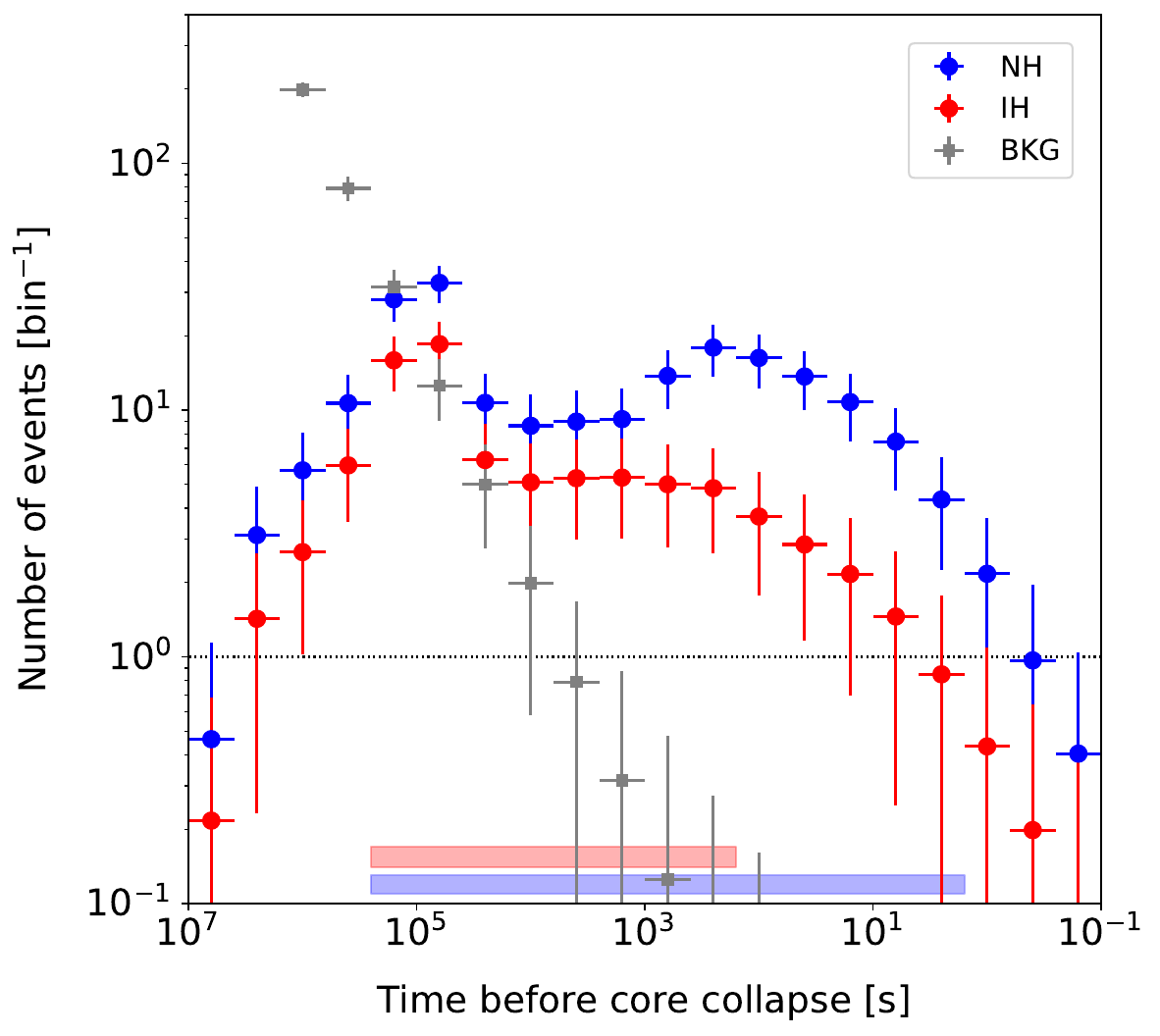}{0.45\textwidth}{(a--1) pre-SN $\nu$, 300~pc, JUNO}\hspace{-15truept}
            \fig{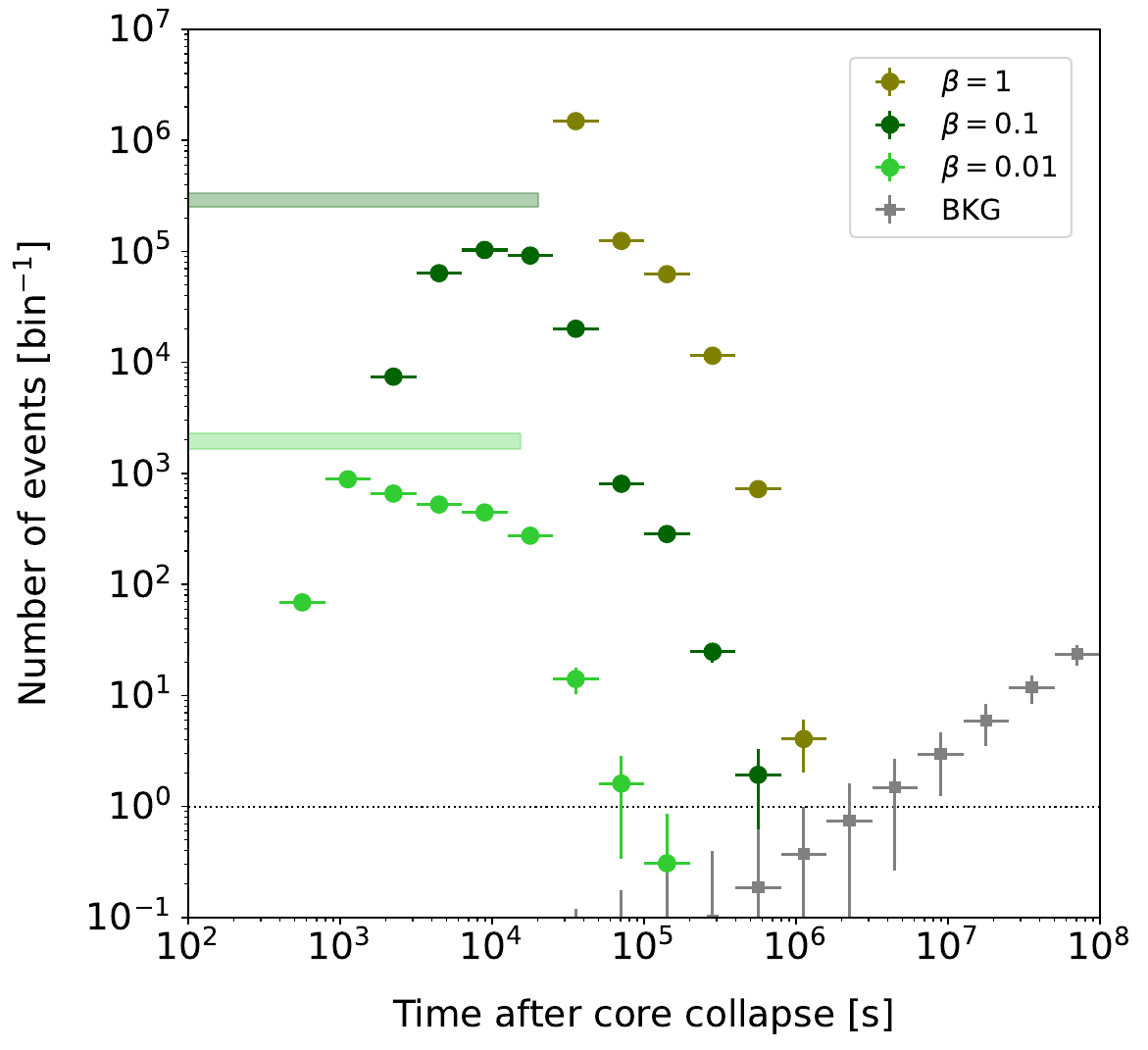}{0.45\textwidth}{(b--1) CSM $\nu$, 300~pc, IceCube}}
    \gridline{\fig{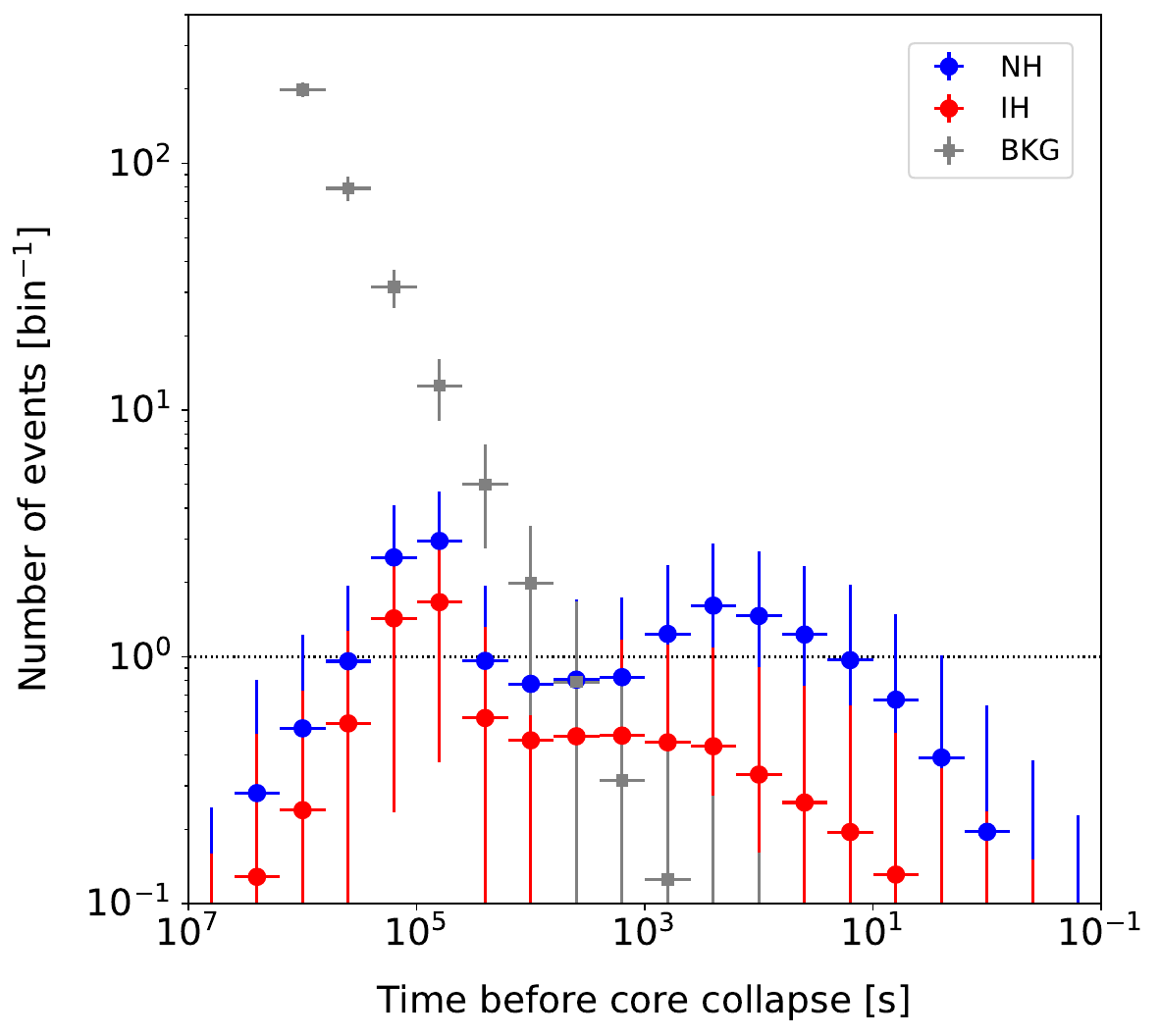}{0.45\textwidth}{(a--2) pre-SN $\nu$, 1~kpc, JUNO}\hspace{-15truept}
            \fig{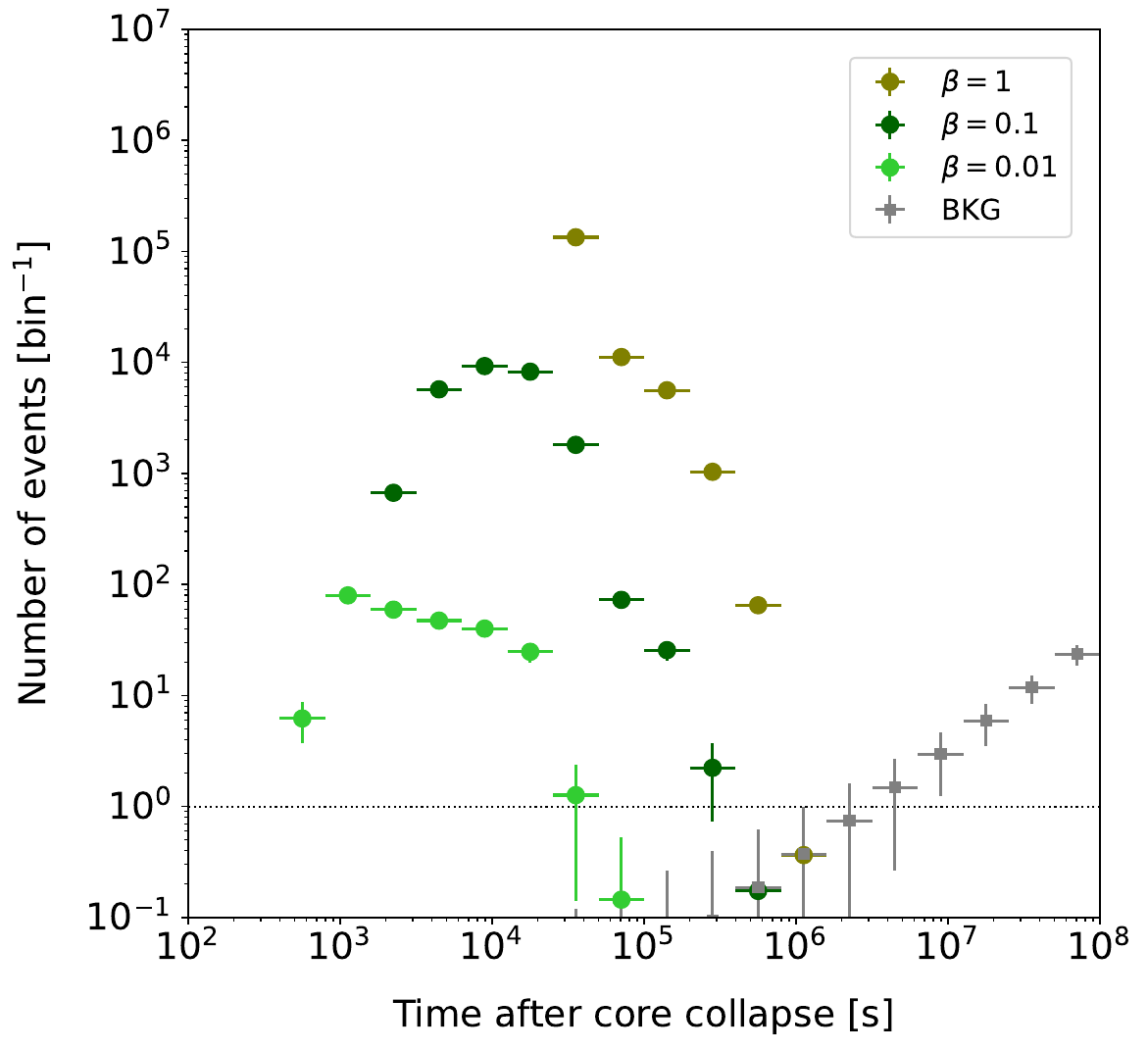}{0.45\textwidth}{(b--2) CSM $\nu$, 1~kpc, IceCube}}
    %\vspace{-10truept}
    \caption{Detected number of signal neutrinos as well as background events along time at (a) JUNO and (b) IceCube from the SN at a distance of 300~pc (top) and 1~kpc (bottom). In the panels (a), the detectable time ranges based on the criteria described in the main text in the case of NH and IH are shown with the blue and red bars, respectively. In the panels (b), the corresponding time range in the case of NH is shown for each mass-loss efficiency case with the different colored bars. The result in the IH case is given in Appendix~\ref{sec:otherresult}.}
\label{fig:eventhist}
\end{figure*}

\begin{figure}[htbp]
    \centering
    \includegraphics[width=0.4\textwidth]{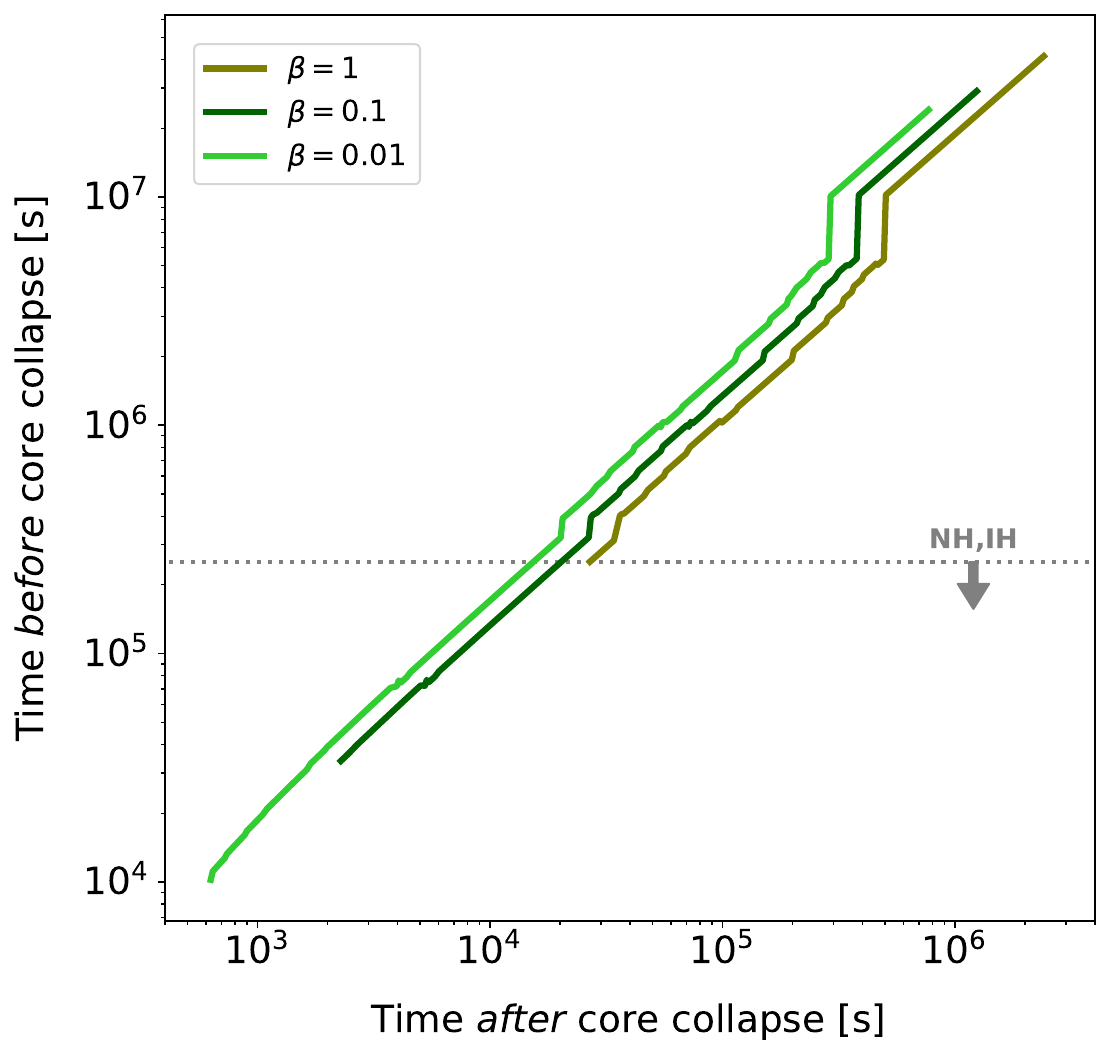}
    %\vspace{+5truept}
    \caption{Correlation between the times before and after core collapse of the progenitor star at three different mass-loss efficiencies. The gray line indicates the detectable time range at JUNO for a 300~pc away SN in the NH and IH cases.}
\label{fig:tcorr}
\end{figure}

\begin{figure}[htbp]
    \centering
    \includegraphics[width=0.48\textwidth]{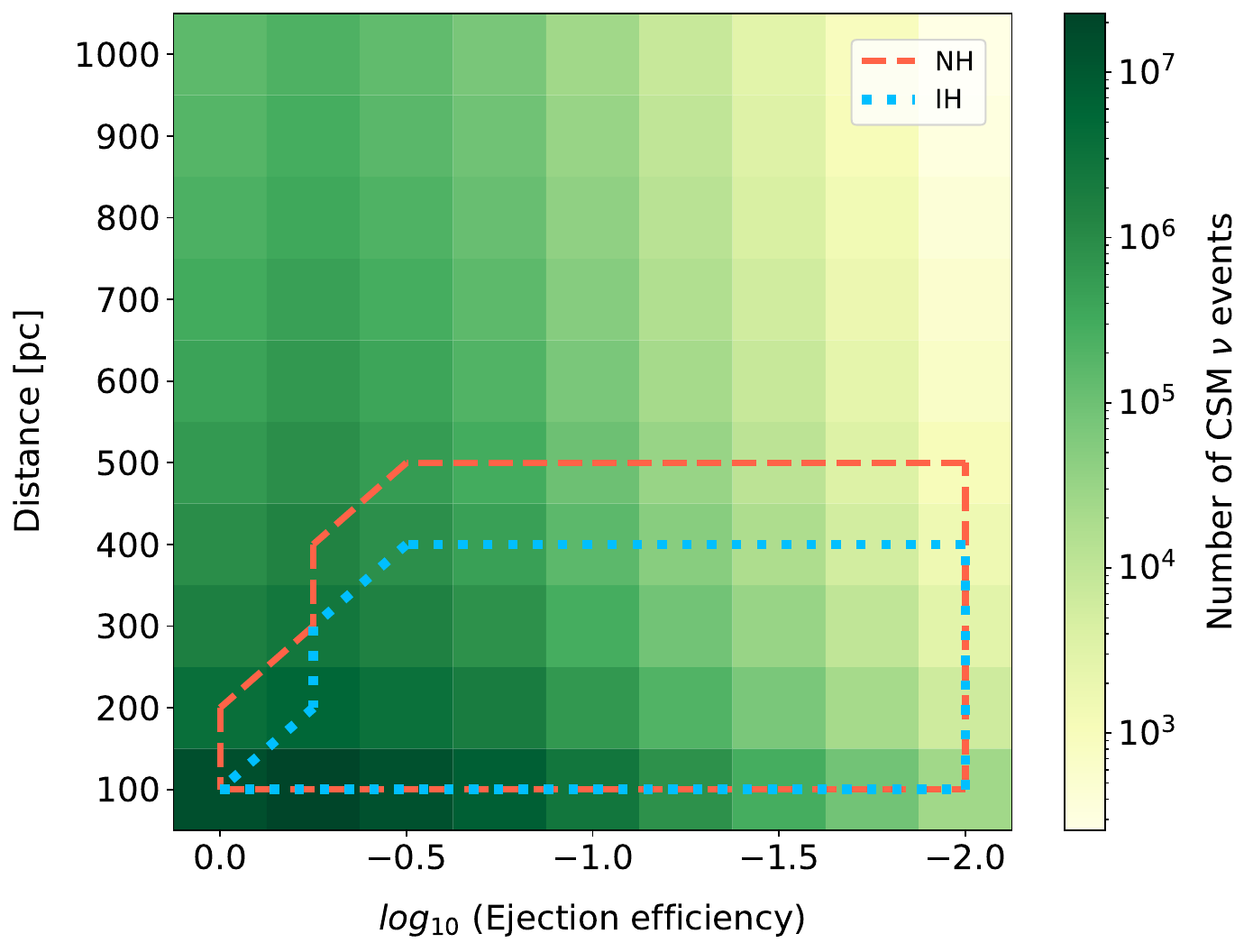}
    %\vspace{+5truept}
    \caption{Parameter space on distance ($d$) and mass-loss efficiency ($\beta$) in which the proposed diagnosis method is applicable with the presented choice on models and detectors, for both NH and IH cases, with overlaid on the number of CSM neutrino events expected at IceCube.}
\label{fig:appspace}
\end{figure}

%When we return to our strategy in Figure \ref{fig:flow}, in actual statistical tests, we can only use the time evolution of CSM neutrino events reconstructed from the time range in which pre-SN neutrino events were observed.
Now returning to the strategy outlined in Figure~\ref{fig:flow}, we are able to use, in the statistical test (e.g., Kolmogorov–Smirnov test), only the time evolution of the detected CSM neutrinos that fall into the time range in which pre-SN neutrinos are correspondingly observed. 
Pre-SN neutrino detectable time ranges in the left panels of Figure~\ref{fig:eventhist} are converted to the time after core collapse using the relations in Figure~\ref{fig:tcorr}, and the NH cases are shown in the right panels of Figure~\ref{fig:eventhist}. 
%In the case of 300~pc, any choice of mass-loss efficiency can be tested with enough statistics, but it is not likely possibly for the 1~kpc case even though the statistical amount of detected events at IceCube is still large. 
In the case of 300~pc, the mass-loss efficiency choice of 0.1 and 0.01 can be tested with enough statistics, but it is not likely possible for the 1~kpc case even though the statistical amount of detected events at IceCube is still large.
This is because the time range for which synchronized detections of pre-SN and CSM neutrinos are made is determined mainly by the pre-SN neutrino side. 
By repeating the same procedure described above, the applicable parameter space on ($d, \beta$) is drawn in Figure~\ref{fig:appspace}. 
It is interestingly found that the largest mass-loss efficiency case $(\beta=1)$ is relatively harder to test even at closer distances because neutrino emission begins at later times which may be out of the detectable range for pre-SN neutrinos.
%From these results, we claim that the presented diagnosis method is available up to $\sim$1~kpc with some possible improvements as described in Sec.~\ref{sec:concl}. 
The proposed method is available up to $\sim$500~pc with the current choice of model and detector setup. 
This can be improved by many factors described in Sec.~\ref{sec:concl} up to $\sim$1~kpc and even farther in the future.

%<<<
%>>> 

%<<< 
%>>> 

%<<< 
%>>> 

%%---------------------------------------------------------------------------------
\section{Conclusion and Discussion } \label{sec:concl}

%%% summary from the main part 
In this study, we pursue a potential use of neutrinos at different energies to investigate the origin of enhanced CSM around stellar core collapse, and propose a strategy to test whether pre-SN neutrino release is responsible for the enhanced CSM (Figure \ref{fig:flow}). 
Referring to a recent calculation of pre-SN neutrino emission from a 15$M_\odot$ progenitor~\citep[][]{2017ApJ...848...48K}, we reconstruct the CSM density profile that leads to non-thermal high-energy neutrino emission. 
The reconstructed high-energy neutrino light curve is found to reflect the original pre-SN neutrino light curve, and our method is flexibly applicable to any pre-SN model. 
The proposed idea is demonstrated by estimating the event rate of pre-SN neutrinos at JUNO and of CSM neutrinos at IceCube as representative detectors in the MeV and TeV energy regions, respectively. 
The current choice of signal and background models allows us to use the presented diagnosis method for the SN up to about 1~kpc, while considering some uncertain factors as pointed out below.

%%% experimental summary & discussion points 
The presented method is, as shown in Figure~\ref{fig:appspace}, well available up to $\sim$500~pc in NH, and needs some effort to be used at farther distances or in IH. 
There are several possibilities to extend the applicable range.
First, in Sec.~\ref{sec:demo}, we limited the analysis energy region to only 1.8--4.0~MeV as reliable background estimation is provided for this range; however, expanding the analysis window would easily help to gain statistics by up to a factor of a few. 
Second, further background reduction may be possible, in particular, reactor neutrinos may get decreased depending on reactor operation status.  
Third, the light curve analysis could be improved from the current method based on a simple event counting.
Fourth, we would be able to utilize other low-energy neutrino detectors. 
For example, Super-Kamiokande had been upgraded with a gadolinium loading as SK-Gd~\citep[][]{2022NIMPA102766248A,2024NIMPA106569480A}, mainly for the purpose of the diffuse supernova neutrino background search~\citep[][]{2004PhRvL..93q1101B,2023ApJ...951L..27H}, and is now more capable of catching neutrinos around energies of a few to 10~MeV~\citep[][]{2019ApJ...885..133S}. 
Similarly, there is a potential option of loading gadolinium to Hyper-Kamiokande. 
In addition, for the further future, some water-based liquid scintillator R\&D projects are running, targeting low-energy neutrinos with a large volume~\citep[e.g.,][]{2011NIMPA.660...51Y,2020EPJC...80..416A,2024JInst..19P7014Z}.
Combining data from different detectors would add more pre-SN neutrino signals in our scheme and therefore extend the time range for which the synchronized detection with high-energy neutrinos is achieved, realizing roughly a few kpc reach of the CSM diagnosis application. 
It should also be noted that the applicable range mentioned in the main part is tied with our model choice~\citep[][]{2017ApJ...848...48K} and hence can be longer or shorter in reality. 

The application range of the current method is limited mainly by pre-SN neutrino detection, but one can detect CSM neutrinos alone at IceCube and KM3NeT farther away~\citep[][]{2023ApJ...945...98V,2023ApJ...956L...8K}, or at a smaller detector, such as SK-Gd and Hyper-Kamiokande. 
A rough estimation assuming the effective detector volume of SK-Gd (Hyper-Kamiokande) is $10^{-4}$ ($10^{-3}$) times smaller than IceCube still gives $\mathcal{O}(10)$ ($\mathcal{O}(100)$) events from the SN at a distance of 1~kpc and with a mass-loss efficiency of 0.1 (see Figures~\ref{fig:eventhist} and \ref{fig:appspace}). 
Even if considering more realistic situations at these detectors, signals from such CSM-origin neutrinos are expected to be feasibly detected. 

The main scope of this paper is to show the potentiality that we could reject or accept one hypothesis about the origin of the CSM when we observe enough signals in two energy regions; however, it may still be useful to raise potential systematic factors that can affect our estimation. 
The uncertainty of our model includes the SN profiles, such as explosion energy and ejecta mass, but they only change the scale of the observed number of signals. 
If the CSM origin is the release of pre-SN neutrinos, we can still capture the correlation in time structure that would not appear in other CSM origins. 
Another possible uncertainty comes from the experimental estimation and is actually dominated by data statistics.
Nevertheless, at 300 pc, as shown in Figure~\ref{fig:eventhist}, the time structure can be captured very well with the considered uncertainty.

%%% theoretical summary & discussion points 
It should be noted that the population of known Wolf-Rayet (WR) stars, which is the progenitor of stripped supernovae, within $\sim$1 kpc of the Earth is not large:
the closest confirmed WR star is $\gamma^2$ Velorum (WR 11) at $\sim$340~pc \citep[][]{2007MNRAS.377..415N}, and only a few have been firmly identified within $\sim$1~kpc \citep[][]{1988A&A...199..217V}. 
While recognizing that the number of available targets is limited, the case of 300 pc is designed for these nearby astronomical objects. 
Moreover, with future improvements in observation capabilities extended to a few kpc, we expect that the number of detectable WR stars will increase to more than 10.

%%% last words 
The presented work leaves a first footprint in the field of {\it multi energy neutrino astronomy} towards the era of various large-scale neutrino telescopes, and may stimulate studies of relevant topics.
In compound with multi-messenger astronomy, these studies would help to explore stars and the Universe more deeply.

%Cite a paper ``Low- and High-energy Neutrinos from SN 2023ixf in M101'' \citep[][]{2023ApJ...955L...9G} somewhere?

%%---------------------------------------------------------------------------------
\section*{Acknowledgments}
The authors thank Yudai Suwa, Kohta Murase and Chinami Kato for their helpful comments and suggestions.
%The authors are also grateful to the anonymous referee for his/her comments improving our manuscript.
Discussions during the YITP workshop YITP-W-24-22 on ``Exploring Extreme Transients: Emerging Frontiers and Challenges'' were also useful to complete this work.
This work has been supported by the Japan Society for the Promotion of Science (JSPS) KAKENHI grants 21K13964 and 22KJ0528 (RS), and JSPS KAKENHI grant 22H04943 and National Science Foundation (NSF) grant PHY-2309967 (YA).

%%---------------------------------------------------------------------------------
%\clearpage
\restartappendixnumbering
\appendix

%%% 
\section{SN ejecta profile} \label{app:sn}
In order to solve Eqs. \eqref{eq:conv1} and \eqref{eq:conv2}, we assume that the unshocked SN ejecta expand homologously $(r=v\cdot t)$.
We further approximate its density profile $\rho_\mathrm{SN}(r)$ by a broken power-law \citep[e.g.,][]{1994ApJ...420..268C,1999ApJ...510..379M}:
\begin{equation}
  \rho_\mathrm{SN}(r)= \left\{
  \begin{aligned} 
    \rho_0 (v/v_\star)^{-\delta} \qquad \mathrm{at}\quad v \ge v_\star\\
    \rho_0 (v/v_\star)^{-n} \qquad \mathrm{at}\quad v < v_\star
  \end{aligned}
  \right. ~,\label{eq:snprofile1}
\end{equation}
where the normalization $\rho_0$ is chosen so that integrating $\rho_\mathrm{SN}(r)$ over velocity space yields the total ejecta mass $M_\mathrm{SN}$. 
The characteristic velocity $v_\star$ is given by
\begin{equation}
  v_\star = \sqrt{\cfrac{2(5-\delta)(n-5)E_\mathrm{SN}}{(3-\delta)(n-3)M_\mathrm{SN}}} ~,\label{eq:snprofile2}
\end{equation}
where $E_\mathrm{SN}$ is the kinetic energy of the SN ejecta. Typically, $\delta$ is from 0 to 1. 
For red supergiant progenitors, 
$n \approx 12$ is expected, whereas $n \approx 10$ is more typical of stars with radiative envelopes for progenitors with radiative envelopes \citep[e.g., blue supergiants;][]{1999ApJ...510..379M}. 
Hence, the density $\rho_\mathrm{ej}$ of the unshocked SN ejecta at $R_\mathrm{sh}$ is given by $\rho_\mathrm{ej}=\rho_\mathrm{SN}(v_\mathrm{ej})$.

%%% 
\section{Other Setup Results} \label{sec:otherresult}

\begin{figure*}[htbp]
    \gridline{\fig{nevent_binned_presn_m15_juno_300pc.pdf}{0.45\textwidth}{(a) 300~pc}\hspace{-15truept}
            \fig{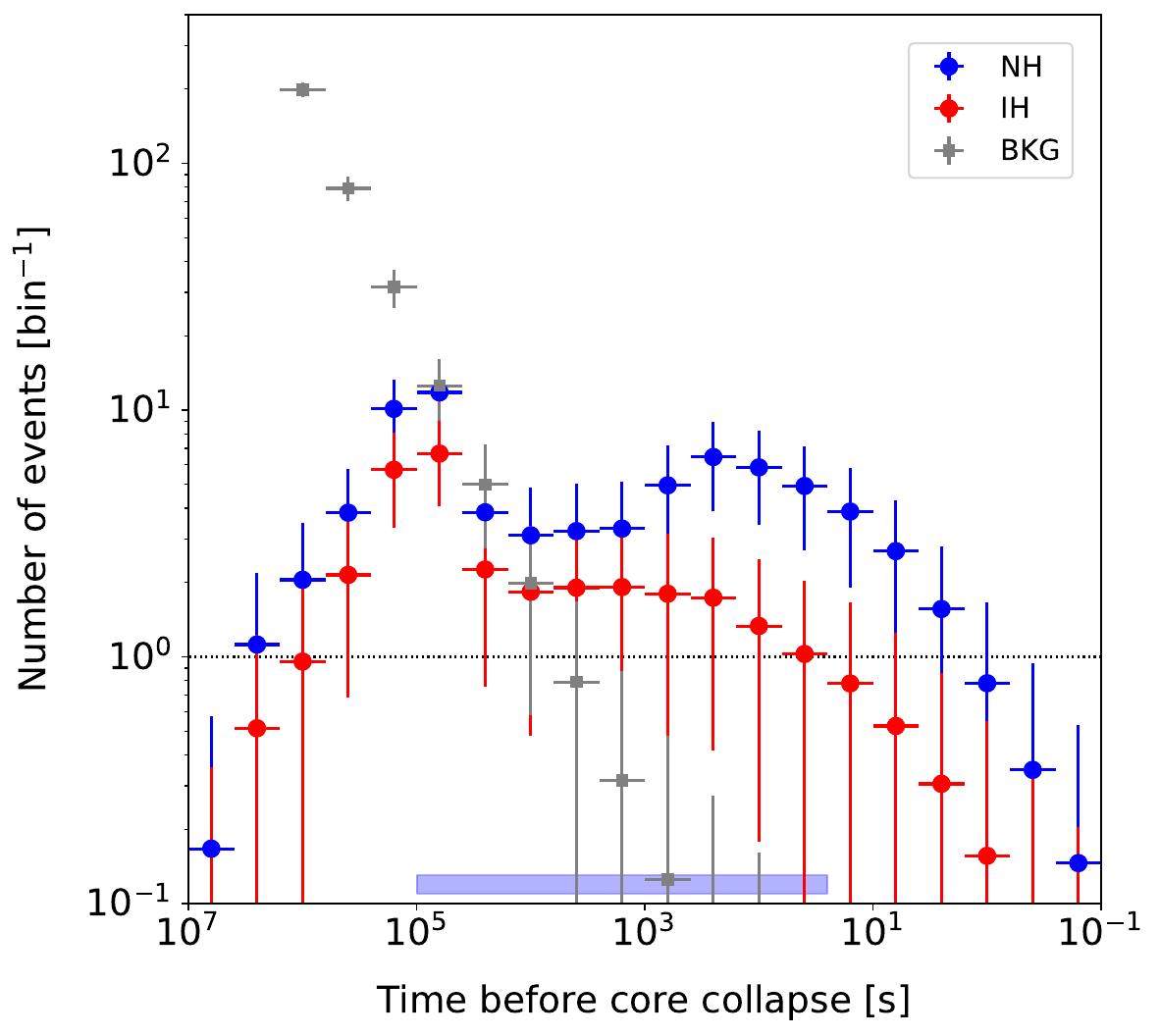}{0.45\textwidth}{(b) 500~pc}}
    \gridline{\fig{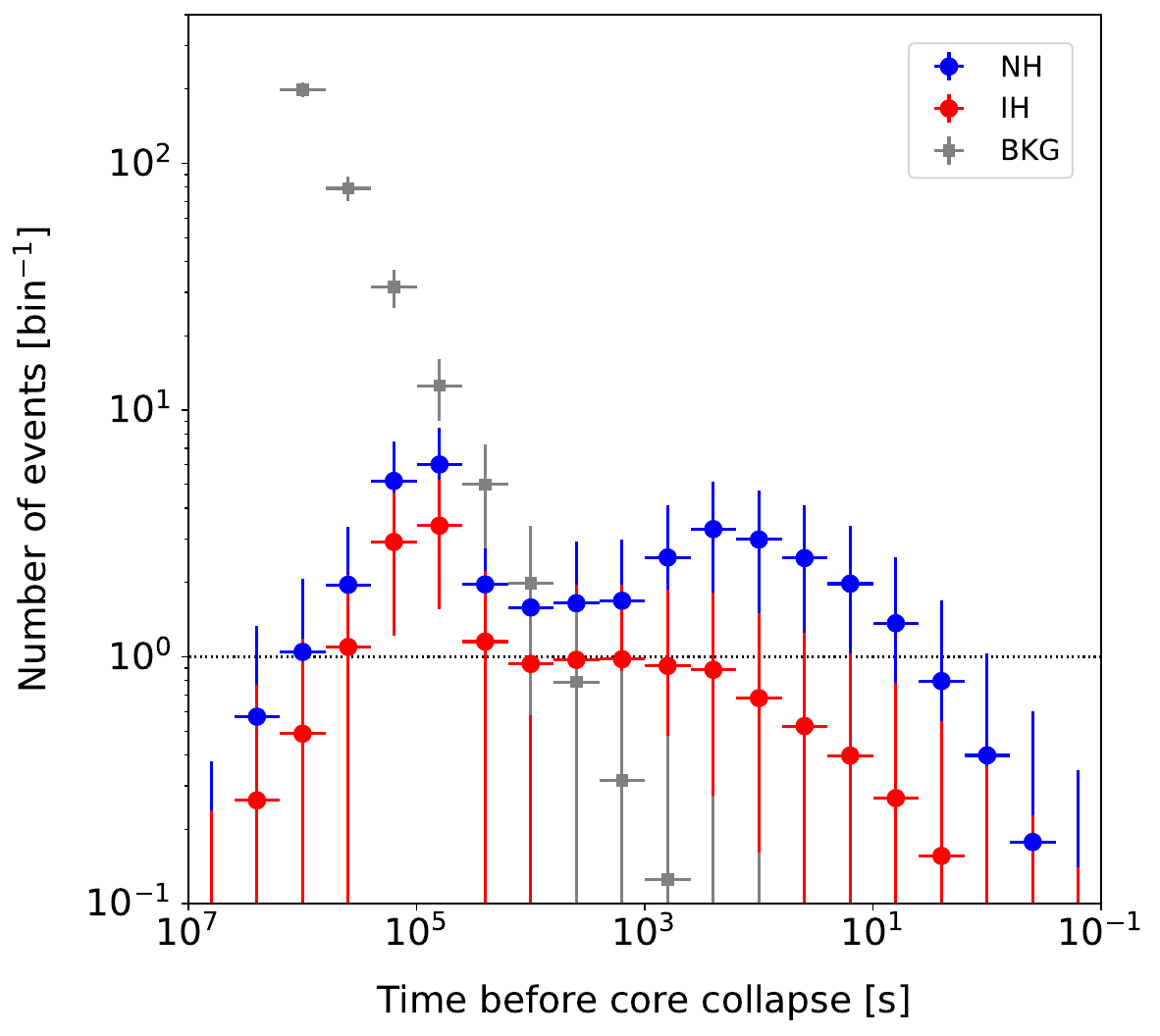}{0.45\textwidth}{(c) 700~pc}\hspace{-15truept}
            \fig{nevent_binned_presn_m15_juno_1kpc.pdf}{0.45\textwidth}{(d) 1~kpc}}
    %\vspace{-10truept}
    \caption{Detected number of pre-SN neutrinos as well as background events along time at JUNO from the SN at a distance of (a) 300~pc, (b) 500~pc, (c) 700~pc, and (d) 1~kpc. The color notations are the same as Figure~\ref{fig:eventhist}.}
\label{fig:otherpresneventhist}
\end{figure*}

\begin{figure*}[htbp]
    \gridline{\fig{nevent_binned_csm_m15_icecube_300pc_nh.pdf}{0.45\textwidth}{(a--1) 300~pc, NH}\hspace{-15truept}
            \fig{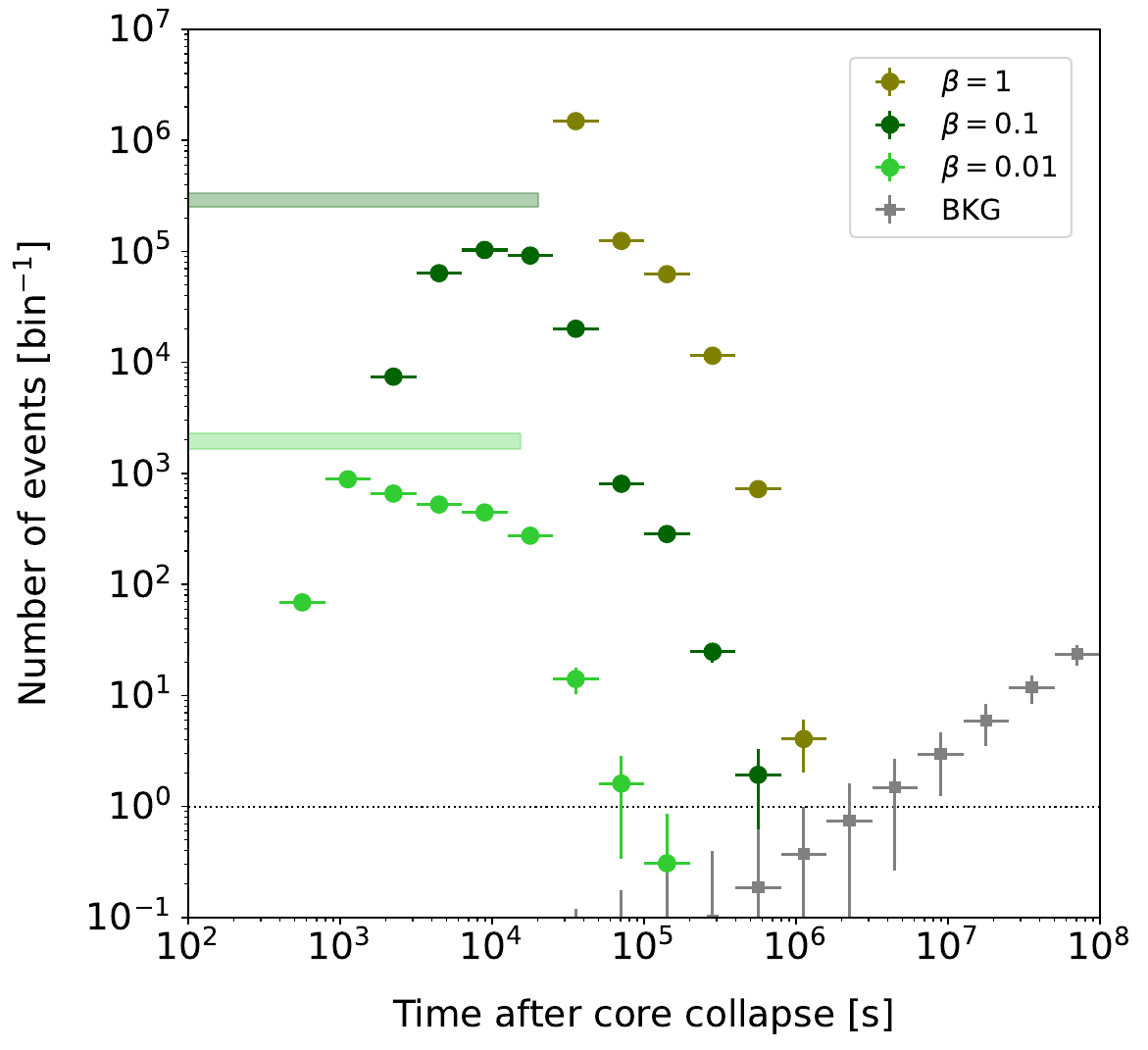}{0.45\textwidth}{(b--1) 300~pc, IH}}
    \gridline{\fig{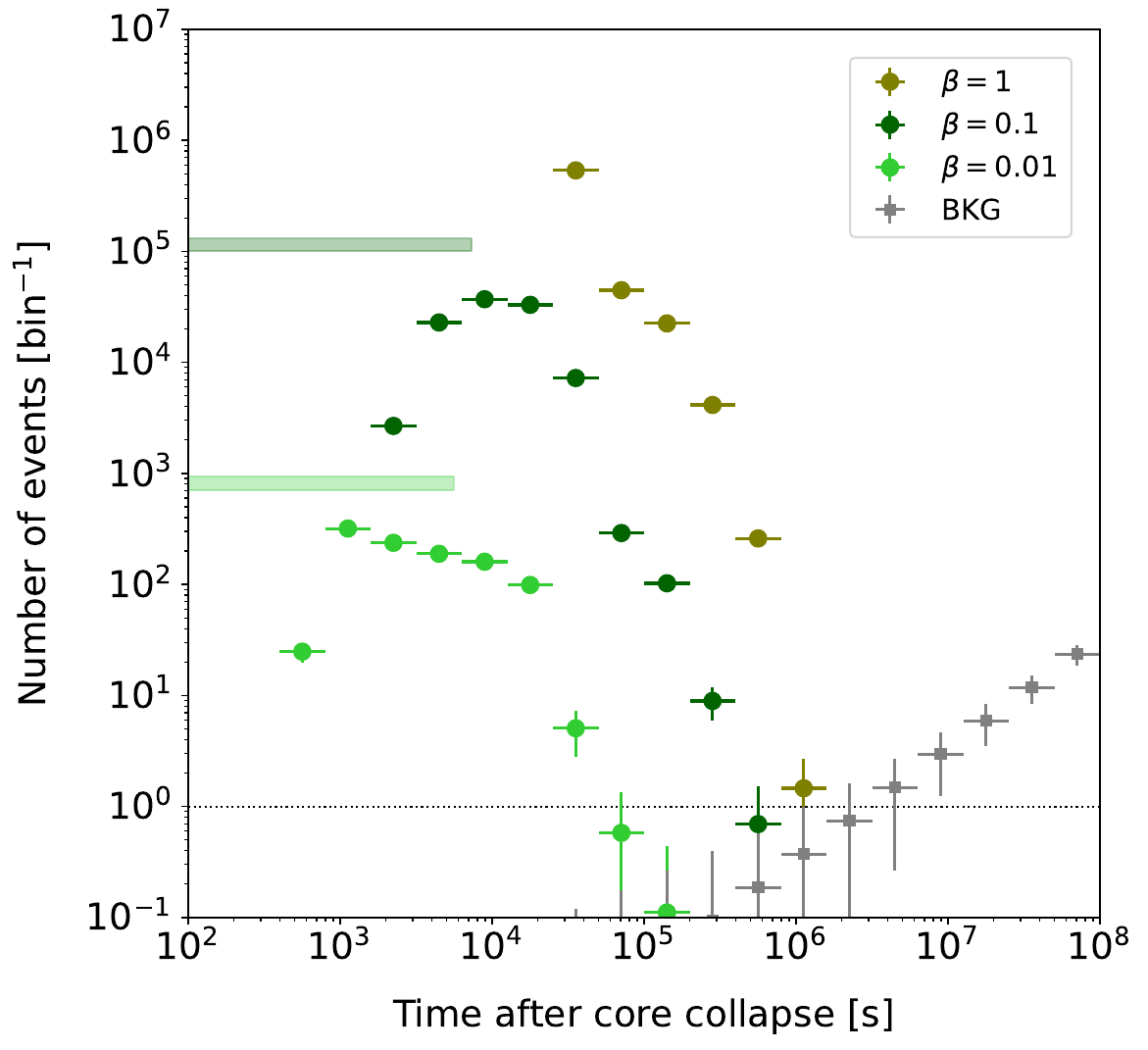}{0.45\textwidth}{(a--2) 500~pc, NH}\hspace{-15truept}
            \fig{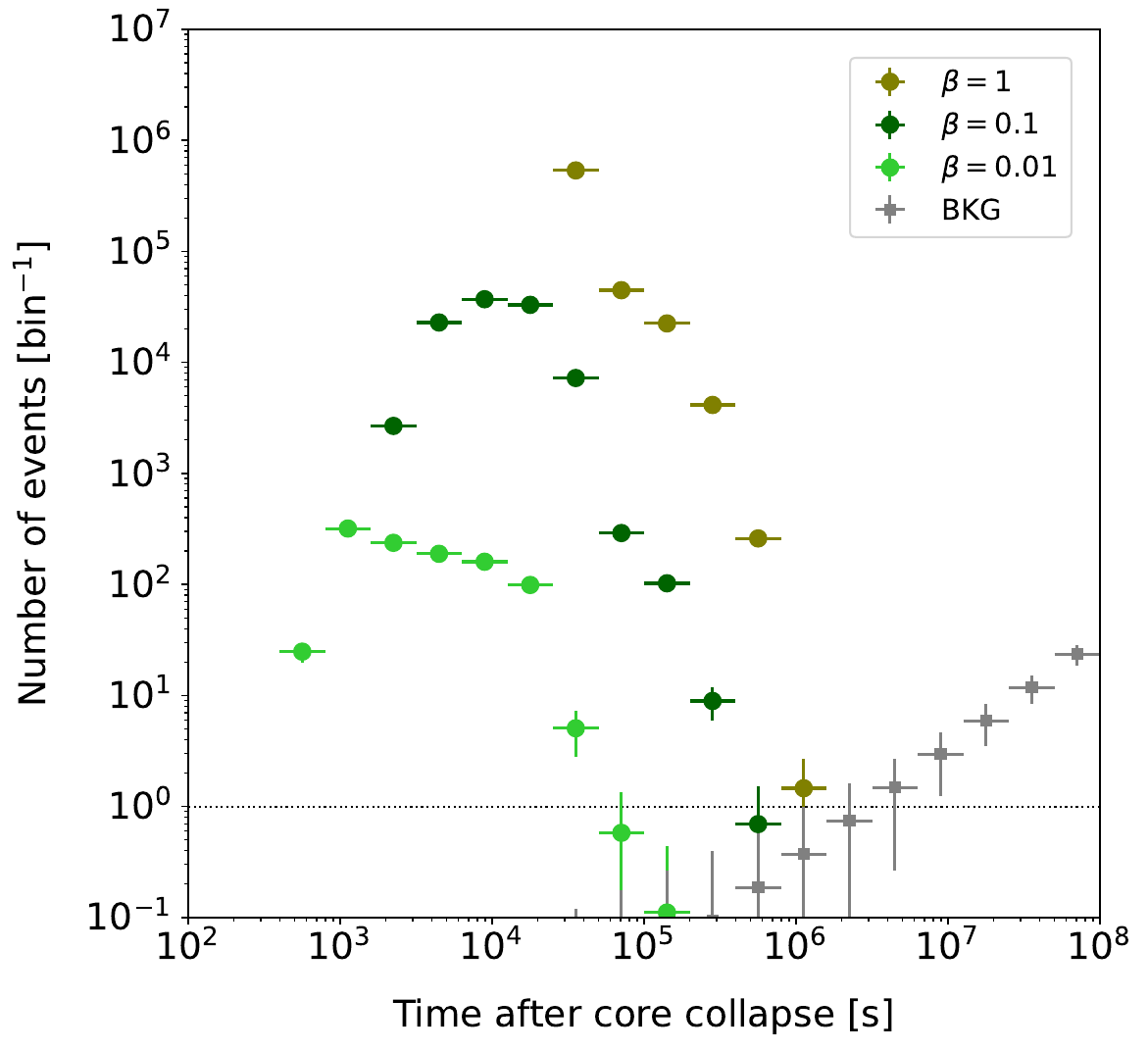}{0.45\textwidth}{(b--2) 500~pc, IH}}
    \caption{Detected number of CSM neutrinos as well as background events along time at IceCube from the SN at a distance of (1) 300~pc and (2) 500~pc for the case of (a) NH and (b) IH in the pre-SN neutrino sector. The color notations are the same as Figure~\ref{fig:eventhist}.}
\label{fig:othercsmeventhist}
\end{figure*}

Figure~\ref{fig:otherpresneventhist} shows the pre-SN neutrino event rate at JUNO for the different distances, and Figure~\ref{fig:othercsmeventhist} shows the CSM neutrino event rate at IceCube for the different distances and neutrino mass hierarchy cases. 
The used models and methods are outlined in the main part of the paper. 

%<<<
%>>> 

%<<<
%>>>

%\clearpage
%\newpage
\bibliography{ref}{}
\bibliographystyle{aasjournal}



\end{document}